\def\tr{{\rm tr}\,}
\def\Tr{{\rm Tr}\,}

\def\b{\bibitem}
\documentstyle[aps,pra,eqsecnum,psfig,floats]{revtex} 
\begin{document}
\def\SNG{{\em Physical Review Style and Notation Guide}}
\def\LUG {{\em \LaTeX{} User's Guide \& Reference Manual}}
\def\btt#1{{\tt$\backslash$\string#1}}%
\def\REVTeX{REV\TeX}
\def\AmS{{\protect\the\textfont2
        A\kern-.1667em\lower.5ex\hbox{M}\kern-.125emS}}
\def\AmSLaTeX{\AmS-\LaTeX}
\def\BibTeX{\rm B{\sc ib}\TeX}
\title{Quantum phase transitions in electronic systems}
        
\author{T.R.Kirkpatrick}
\address{Institute for Physical Science and Technology, and Department of 
         Physics\\
         University of Maryland, College Park, MD 20742}
\author{D.Belitz}
\address{Department of Physics and Materials Science Institute,
         University of Oregon,
         Eugene, OR 97403}
\date{\today}
\maketitle
\begin{abstract}
Zero--temperature or quantum phase transitions in itinerant electronic
systems both with and without quenched disordered are discussed. Phase
transitions considered include, the
ferromagnetic transition, the antiferromagnetic transition, the
superconductor--metal transition, and various metal--insulator transitions. 
Emphasis is placed on how to determine the
universal properties that characterize these quantum phase transitions.
For the first three of the phase transitions listed above, one
of the main physical ideas established is that in zero--temperature systems
there are soft or slow modes that exist in addition to the soft order
parameter fluctuations, and that these modes can
couple to the critical modes. These extra soft modes are shown to have a
profound effect on the quantum critical properties. For quantum phase
transitions involving zero wavenumber order parameters, i.e., the
ferromagnetic and superconductor--metal transitions, these extra modes
effectively lead to long--ranged effective interactions between order
parameter fluctuations, which in turn lead to exactly
soluble critical behaviors. For the antiferromagnetic case, 
we argue that while in low
enough dimensions disorder fluctuation effects tend to destroy long--range 
order, quantum fluctuations counteract this effect and in some parameter
regions manage to re--establish antiferromagnetic long--range order.
For the metal--insulator transition, some recent new ideas are reviewed.
In particular, it is pointed out that for interacting disordered electrons,
one expects that in high dimensions the metal--insulator transition is 
related to the phase
transition that occurs in random--field magnets in high dimensions. If the
analogy also holds in three dimensions this suggests that the  
metal--insulator transition might have glassy characteristics.
\vskip 90mm
\par\noindent
Contribution to {\it Electron Correlation in the Solid State}, Norman H. March
(editor) 
\par\noindent
To be published by Imperial College Press/World Scientific
\end{abstract}

\vfill\eject
\mbox{}
\vskip 20mm
\tableofcontents
\vskip 20mm
\par\noindent
\centerline{\bf Abbreviations}
\bigskip
\begin{tabbing}
AFM\qquad  \= Antiferromagnetism\\
\medskip
LGW \> Landau--Ginzburg--Wilson\\
\medskip
MIT \> Metal--insulator transition\\
\medskip
QPT \> Quantum Phase Transition\\
\end{tabbing}
\vfill\eject

\section{Introduction}
\label{sec:I}

The study of continuous phase transitions
\footnote[1]{We will deal only with
continuous transitions, and will simply refer to them as `phase transitions' or
`critical points'.} 
has led to great advances in 
condensed matter physics. The insights gained from these 
studies\cite{PhaseTransitions} have had
remarkably broad effects, which is somewhat suprising given that 
phase transitions occur only at very special points in
the phase diagram. In particular, phase transition theory spawned
the application of renormalization techniques in condensed matter
physics\cite{Wilson}, which have proved to be very powerful in a more general
context. Wilson's breakthrough paved the way for the 
understanding of thermal or classical
phase transitions, which occur at non--zero temperature and are driven
by thermal fluctuations. Let the critical temperature be $T_c>0$, and
denote the distance from the critical point by $t = 1 - T/T_c$. Then
it turns out that the critical behavior asymptotically close to the
critical point is entirely determined by classical physics. This can
be seen as follows. Upon approaching the critical point, there is a
diverging characteristic length in the system, viz. the correlation
length $\xi\sim \vert t\vert^{-\nu}$, with $\nu$ an exponent that is
characteristic of the class of phase transitions under consideration.
Also, the characteristic time $\xi_{\tau}$ for relaxation towards 
equilibrium diverges like $\xi_{\tau}\sim\xi^z$, with $z$ another
exponent. This slow relaxation is caused by slow fluctuations of some
thermodynamic quantity, the order parameter, that is characteristic of
the transition. These fluctuations relax on microscopic time scales
far from the transition, but become infinitely slow as $t\rightarrow 0$.
This means there is a characteristic frequency, 
$\omega_c\sim 1/\xi_{\tau}$, that vanishes at criticality. Now quantum
effects will be unimportant as long as $\hbar\omega_c << k_B T$, which
is always the case sufficiently close to a critical point with a
non--zero $T_c$. As a result, the critical behavior at, say, the
Curie point in Iron, the $\lambda$--transition in liquid Helium,
or the superconducting transition in Mercury, are determined
entirely by classical physics,
notwithstanding the fact that quantum mechanics is
crucial for the microscopic mechanism underlying the transition
and, in the last two examples, for the properties of the ordered phase.

\begin{figure}[thb]
\centerline{\psfig{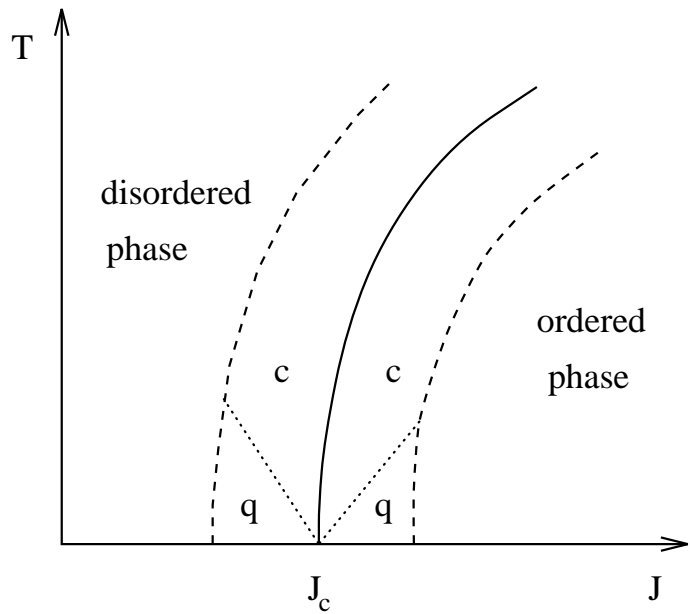}\vspace*{5mm}}
\caption{Schematic phase diagram with $J$ some parameter that is responsible
for the ordering. The quantum critical point is at $(T=0,J_c)$. The dashed
lines denote the boundaries of the critical region, and the dotted lines
mark the crossover from quantum to classical critical behavior. The regions
labeled by c and q show classical and quantum critical behavior,
respectively.}
\label{fig:1.1}
\bigskip
\end{figure}

These considerations raise the interesting question of what happens
when we follow a phase separation line in a phase diagram down to
zero temperature. For definiteness, let us consider a ferromagnet, or a
superconductor, where we decrease $T_c$ by diluting it with a non--magnetic
or non--superconducting material. Then we obtain a schematic phase
diagram like the one shown in Fig.\ \ref{fig:1.1}, with $J$ some parameter
that triggers the transition.
Normally, $J$ is fixed for a given material, and one observes the transition
by lowering the temperature. However, if we imagine crossing the phase
separation line at $T=0$ by varying $J$, then the above arguments suggest
that quantum effects should be important for the transition, since
$T_c =0$. We therefore expect a critical behavior in this case that is
different from the one observed when crossing the phase separation line
at any non--zero $T$. This question is not academic, as was first pointed
out by Suzuki\cite{Suzuki} and Hertz\cite{Hertz}. If one triggers the
transition at low but non--zero $T$, then one observes a `crossover' from
a region that is characterized by quantum critical behavior to one showing
classical critical behavior, as indicated in the figure. Moreover, there
are phase transitions that occur {\em only} at $T=0$, the most important
examples being various metal--to--insulator transitions, which are
studied experimentally by observing a smeared transition at very 
low temperatures.

Such phase transitions that occur in quantum mechanical systems at
$T=0$ as a function of some non--thermal control parameter are
called quantum phase transitions (QPTs). Like their finite--temperature
counterparts, which for distinction are often referred to as thermal or 
classical phase transitions, they are characterized by a diverging
correlation length $\xi$, and a diverging relaxation time $\xi_{\tau}$.
However, the critical fluctuations that lead to these diverging length and
time scales are quantum fluctuations rather than thermal ones. Apart
from the origin of the relevant fluctuations being different, quantum phase
transitions also differ from their classical counterparts in that
their dynamic and static critical behaviors are coupled together.
Technically this implies that one cannot solve for the thermodynamic
properties of the phase transition without also solving for the dynamics.

There are a number of ways to see the coupling between statics and dynamics
at QPTs. From a general scaling viewpoint one argument is as follows. 
It is obvious from Fig.\ \ref{fig:1.1} and the accompanying discussion
that increasing the temperature from zero takes one away from the quantum
critical point, much like increasing the frequency from zero takes one
away from either a quantum or a classical one. Now the fact that Planck's
constant $\hbar$ and Boltzmann's constant $k_B$ are chosen arbitrarily
implies that in fundamental units,
frequency, energy, and temperature all have the same dimensions.
\footnote[2]{This has recently been emphasized in 
Ref.\ \onlinecite{DamleSachdev}.}
From a scaling
viewpoint, this implies, for example, that if a system is at a quantum
critical point, either going to a finite frequency or increasing the
temperature will affect the system in the same way, and be characterized by
the dynamical scaling exponent that describes the divergence of $\xi_{\tau}$
and is usually denoted by $z$. Since the
thermodynamic observables certainly depend on the temperature, this further
implies that the scaling relations or homogeneity laws for these quantities
will also depend on $z$. That is, the static and
dynamic critical behaviors are coupled together. A more mathematical
argument is that the standard Matsubara frequency technique implies that 
frequency
and temperature are intimately related and should scale in the same way.

From a more technical point of view, a classical partition function
involves separate integrals over momenta and positions that are decoupled
since the kinetic energy depends only on the momenta, and the potential
energy only on the positions. One can therefore solve for the thermodynamics
without having to deal with the dynamics explicitly. In quantum statistical
mechanics, on the other hand, the partition function of a system with
Hamiltonian $H$ is given as
\footnote[3]{We will denote by $\Tr$ traces over all degrees of freedom,
including continuous ones, and by $\tr$ traces over discrete degrees of
freedom that are not shown explicitly.}
\begin{equation}
Z = \Tr e^{-H/k_BT}\quad.
\label{eq:1.1}
\end{equation}
All methods for calculating the above trace make use of the imaginary
time technique\cite{NegeleOrland} 
in one form or another. The latter is based on the observation
that $e^{-\beta H}$, with $\beta = 1/k_BT$, is formally identical with the
time evolution operator $e^{iHt/\hbar}$, with the time $t$ replaced by the
imaginary quantity $i\hbar\beta$. One then proceeds to divide the imaginary
time interval $[0,\beta]$ into many infinitesimal subintervals for which the
time evolution can be calculated. The coupling between statics and dynamics
then arises from the fact that the Hamiltonian taken at some imaginary time,
$H(\tau)$, does not commute with $H$ taken at another imaginary time $\tau'$.

Although QPTs are currently a topic of great interest in condensed matter
physics, they are actually a subject with a long history. The earliest work
all dealt with quantum spin systems, and especially one--dimensional chain
models. Important work was done by Bethe, Bloch, Tomonaga, and Lieb. In the
mid--1970's the QPT problem was cast into a more general theoretical
framework. Suzuki\cite{Suzuki} used the so--called Trotter formula, 
which is one way to
rewrite the trace in Eq.\ (\ref{eq:1.1}) by dividing $[0,\beta]$ into
sub--intervals, to map some $d$--dimensional quantum spin problems onto 
$(d+1)$--dimensional classical statistical mechanical problems. The extra 
dimension was for imaginary time, which is of infinite extent in the 
zero--temperature limit ($\beta\rightarrow \infty$), effectively increasing 
the dimension of the system. Beal--Monod\cite{BealMonod} noted that this 
mapping made it obvious how to generalize the renormalization group techniques 
that were being used to describe classical phase transitions for an 
application to quantum transitions.
Motivated by these works, Hertz\cite{Hertz} made a number of important advances.
First he showed how to derive Landau--Ginzburg--Wilson (LGW) order--parameter
functionals, starting from a fermionic description of itinerant electron
systems that writes the trace in Eq.\ (\ref{eq:1.1}) as a functional integral. 
As expected, he showed that to describe QPTs one needs to
consistently describe both space and (imaginary) time fluctuations together.
He further showed that Suzuki's mapping of a $d$--dimensional
quantum system on a $(d+1)$--dimensional classical one was too restrictive for
generic systems. He found that, more generally, a $d$--dimensional quantum 
system is related to
a $(d+z)$--dimensional classical one, with $z$ the dynamical scaling exponent 
for the quantum phase transition. The need for $z\neq 1$ arises since in
general in a condensed matter system, space and time do not scale in the
same way.
%
%
For instance, in a system with diffusive dynamics, one expects
frequency to scale like the wavenumber squared, which suggests $z=2$.
The special case $z=1$ occurs when space and time happen to scale in the 
same way. Hertz then showed that the same renormalization
group ideas that had been used to characterize classical phase transitions
could be applied to the quantum case for arbitrary values of $z$.

Hertz concluded that most QPTs in
three--dimensional ($3$--$d$) systems are trivial from
a phase transition viewpoint in the sense that they are characterized by
Landau or mean--field critical exponents. This conclusion was based on
the observation that the mapping $d\rightarrow d+z$, with $z>1$, lowers
the upper critical dimension $d_c^+$, above which the transition is correctly
described by a mean--field/Gaussian theory, from $d_c^+ = 4$ for most
classical phase transitions to $d_c^+ = 4-z < 3$ for the quantum transition.
For systems with $z=1$ the critical behavior would still be mean--field like
up to logarithmic corrections to scaling. The
only exceptions would be low--dimensional $(d\leq 2)$ systems. 

One of the
main points of the current review is to emphasize that most QPTs are
more complicated than suggested by the above arguments. These complications
arise because in many quantum systems there are soft or slow modes that
exist in addition to the critical order parameter fluctuations. 
These ``extra'' modes arise either
from conservation laws (recall that statics and dynamics are coupled at
QPTs), or from Goldstone modes that result from some spontaneously
broken continuous
symmetry. These modes will in general couple to the critical modes and 
influence the critical behavior. If one insists on integrating out these 
degrees of freedom in order to obtain a standard LGW theory that is formulated
in terms of the order parameter only, then this LGW theory becomes non--local.
Physically, the extra modes lead to long--range interactions
between the order parameter fluctuations, which in turn leads to a
non--local LGW functional. This breakdown of Hertz's theory becomes
particularly obvious in systems with quenched disorder.
In the cases considered by Hertz, he again
found that the correlation length exponent had its mean--field value,
$\nu =1/2$, for all physical values of the dimensionality. However, it is 
now known (though it was not known when Hertz derived his theory) that $\nu$
must obey the inequality $\nu\geq 2/d$ in generic disordered
systems\cite{Harris}. This implies that the mean--field result cannot
be correct for $d<4$.

The LGW theories mentioned above are one example of what is called effective
field theories for condensed matter systems. The main motivation for deriving
and studying such effective field theories is that often one is not
interested in all of the detailed behavior that arises from the microscopic
properties of a system. For instance, one might ask about universal effects
that are the same for large classes of systems, even though the members of
these classes are different with respect to their composition, their
band structure, and other microscopic details. Such universal properties
arise from long--wavelength and low--frequency
fluctuations, i.e. processes that occur on length and time scales that are
large compared to the microscopic lengths and times that determine, say,
the band structure. This in turn implies that only the dominant fluctuations in
this limit need to be retained explicitly
and treated carefully in order to describe the
universal properties. Of particular interest are soft modes, i.e. excitations
that are massless in the limit of zero frequencies and wavenumbers.
All other degrees of freedom can be integrated out,
often in a crude approximation, without affecting the description of the
universal behavior. The resulting theory
for the most important fluctuations in a specified system is called an
effective field theory. This is quite analogous to the situation in
high--energy physics, where one also deals with effective field theories
that describe the low--energy or long--distance behavior of some
underlying microscopic theory. The main difference is that in condensed
matter physics the microscopic theory is known, but too complicated to solve
in general, and the problem consists
of deriving an effective theory that is simple enough to extract the
behavior of interest, while in high--energy physics the effective theory
is known (i.e., it has been guessed) and the problem consists of inferring
from it a more microscopic theory. In high--energy physics one thus proceeds
from low to high energies, while the opposite is true in condensed matter
physics. The most well--known examples of effective
field theories in a condensed matter context are those that are used to 
describe the universal behavior near
critical points. The concept is much more general, however. Even far from
any phase transitions, soft modes will
arise from either the conservation laws, or from the existence of Goldstone
modes in a broken symmetry phase. 
Both of these mechanisms are common in quantum statistical
mechanics, and the resulting soft modes determine the physics at large
distances and low energies. 
In addition to these soft modes, near a phase transition the
critical order parameter fluctuations are also soft. One of our objectives
is to review how to derive effective field theories, both near and far 
from phase transition lines or points.

The current chapter is organized as follows. In Section\ \ref{sec:II} 
we give a brief
discussion of scaling theory near (continuous) quantum phase transitions. We
start Section\ \ref{sec:III} with a general discussion of 
the functional field theoretic
description of itinerant electronic systems. We then discuss how to derive
effective field theories for the relevant degrees of freedom, and show
how to describe soft modes in general disordered electronic systems.
We begin Section\ \ref{sec:IV} by reviewing the ferromagnetic phase
transition in both disordered and clean itinerant electron systems. We will
see that weak localization effects, and their clean
system analogs, determine the universal properties of this phase
transition. We then examine the antiferromagnetic transition in disordered
itinerant systems. The most important result is that quantum fluctuation
effects stabilize antiferromagnetic long--range order in certain parameter
regions. In Section\ \ref{sec:V} we discuss some
recent work on the superconductor--metal transition in bulk ($d=3$)
disordered electron systems. 
In Section\ \ref{sec:VI} we consider the disordered Fermi liquid phase, and
its instability against metal--insulator transitions of the Anderson--Mott
type. First we briefly review the theory based on Finkel'stein's
generalized nonlinear sigma--model, and
then we discuss a recent order--parameter approach to the Anderson--Mott
transition. We discuss similarities to the classical random--field magnet
problem and discuss possible implications for the Anderson--Mott transition in
three dimensions.

There are other interesting and important examples of quantum phase transitions
that we will not discuss, for instance quantum spin chains, quantum spin
glasses, and phase transitions in quantum Hall systems. A brief review of the
latter subject has recently been given by Sondhi et al.\cite{Sondhietal}.

\section{Scaling at quantum critical points}
\label{sec:II}

Scaling ideas have been used with great success in the description of
continuous thermal phase transitions since the 
mid--1960's\cite{PhaseTransitions}. These same ideas,
properly reinterpreted, can be applied to the QPT problem. Since the scaling
description of quantum phase transition has been recently reviewed in a
number of places\cite{Sondhietal,SachdevStatPhys} we will be brief here.
 
In general, the mapping $d\rightarrow d+z$, discussed in 
Sec.\ \ref{sec:I} above, suggests
that the classical scaling relations are valid at the QPT, with the only
change being the above substitution. Although this is the standard lore, and
it is true in a formal sense, we will repeatedly encounter situations where
it is {\em not} valid. When such a breakdown of general scaling happens,
we will mention it explicitly. More specifically, our
main conclusions will be as follows. (1) The classical scaling relations that
are independent of dimensionality are also valid at QPTs. (2) The classical
scaling relations that do depend on $d$ (`hyperscaling relations') 
are formally valid with the
substitution of $d\rightarrow d+z$. However, since many of the important
QPTs are above their upper critical dimensionalities, these scaling relations
in general break down because of the presence of dangerous irrelevant 
variables.
\footnote[4]{The concept of dangerous irrelevant variables is defined at the
end of Sec.\ \ref{sec:II}.}
Further, the situation is much more complicated than in classical systems above
their upper critical dimensionality because, as mentioned in 
Sec.\ \ref{sec:I}, the LGW
functional for describing QPT's is often a non--local one. This feature has
been shown to lead to nontrivial, dimensionality dependent, exponents and
scaling relations, even above the upper critical dimension for the QPT. (3)
There are scaling relationships at QPTs that have no classical analog.
These identities follow from the scaling equivalence of energy, inverse
time, and temperature. They depend on the dimensionality and can also be
invalidated by dangerous irrelevant variables.
 
All scaling theories for continuous phase transitions start with the idea of a
divergent correlation length scale, $\xi$. If $t$ is the dimensionless
distance from the critical point, then the divergence of $\xi$ is
characterized by the critical exponent $\nu$,
\begin{mathletters}
\label{eqs:2.1}
\begin{equation}
\xi \sim \frac 1{\mid t\mid ^\nu }\quad.
\label{eq:2.1a}
\end{equation}
Typically $t$ is defined to be greater than zero in the
disordered phase, and less than zero in the ordered one.
At thermal phase transitions, $t$ is given by the reduced temperature,
and it is very easy to control experimentally.
For QPTs $t$ might represent the distance from a
critical interaction strength, or from a critical impurity concentration.
In the quantum case, $t$ is therefore not so
easily controlled, and to vary $t$ different samples are usually used.
\footnote[5]{An exception is the stress tuning technique, where uniaxial or
hydrostatic stress is applied to a sample that is very close to the transition.
The resulting change in the microscopic structure drives the system across the
phase separation line. See Ref.\ \onlinecite{Rosenbaum} for an application.}
In general this means that experimentally probing the quantum critical
region, with the precision necessary for measuring critical exponents, 
is very difficult. 

As we have mentioned in Sec.\ \ref{sec:I}, in addition to the diverging
correlation length we need to deal with a diverging time scale. As the 
critical point is
approached from, for example, the disordered side, the order that is being
built up decays on a longer and longer time scale. This correlation time
$\xi_{\tau}$ diverges at the transition, and the singularity is characterized 
by the dynamical scaling exponent $z$,
\begin{equation}
\xi _\tau \sim \xi ^z\quad.
\label{eq:2.1b}
\end{equation}
\end{mathletters}
Equations\ (\ref{eqs:2.1}) represent an {\em assumption}, namely that both
$\xi$ and $\xi_{\tau}$ diverge as some power of $t$. There is no principal
reason why the singularity might not be stronger or weaker than that, and we 
will indeed encounter examples of exponential divergencies in 
Secs.\ \ref{sec:V} and \ref{sec:VI} below. For now we will continue,
however, with the discussion of the more conventional power--law scaling
assumption.

Before formalizing the scaling ideas, we will first give two examples of
how scaling (or hyperscaling) is used. First consider the singular 
part of the
free energy density, $f_s$. The dimension of $f_s$ is that of an inverse
volume times an energy. Assuming that $\xi $ and $\xi_{\tau}$ are the only
relevant length and energy
\footnote[6]{We will use units such that $\hbar = 1$ 
in what follows.} 
scales in the problem, this suggests
\begin{mathletters}%
\label{eqs:2.2}
\begin{equation}
f_s\sim \xi ^{-(d+z)}\quad.
\label{eq:2.2a}
\end{equation}
Note that for the classical problem, Eq.\ (\ref{eq:2.2a}) is obtained with
$d+z$ replaced by $d$, since in that case the relevant 
energy scale as $\xi\rightarrow\infty$ is just the critical
temperature, $T_c\neq 0$. Using Eq.\ (\ref{eq:2.1b}) with $\xi_{\tau}\sim 1/T$
yields for the temperature dependence of $f_s$ at $t=0$
\begin{equation}
f_s(t=0,T)\sim T^{1+d/z}\quad.
\label{eq:2.2b}
\end{equation}
\end{mathletters}%
The singular part of the specific heat thus behaves as
\begin{equation}
C\sim T^{d/z}\quad.
\label{eq:2.3}
\end{equation}
That is, by using scaling ideas one can express the critical behavior of $C$
in terms of the dynamical critical exponent.

As another example, consider the
conductivity, $\sigma$. In fundamental units, setting $\hbar =k_B=e^2=1$
(with $e$ the electron charge), the
dimension of $\sigma$ is an inverse length to the power $(d-2)$. At a phase
transition where $\sigma$ vanishes in the `disordered' phase and is nonzero
in the `ordered' phase, i.e., at a metal--insulator
transition,
\footnote[7]{We will see in Sec.\ \ref{sec:VI} in which sense the
metallic phase can be considered the ordered one.
}
this suggests that $\sigma$ vanishes as
\begin{mathletters}
\label{eqs:2.4}
\begin{equation}
\sigma(T=0,t) \sim \xi ^{-(d-2)}\sim t^{\nu (d-2)}\quad.
\label{eq:2.4a}
\end{equation}
We see that the conductivity exponent $s$, defined by $\sigma \sim t^s$, 
is related to the correlation length exponent $\nu$ by
\begin{equation}
s=\nu (d-2)\quad.
\label{eq:2.4b}
\end{equation}
\end{mathletters}%
Equation\ (\ref{eq:2.4b}) is called Wegner's scaling relation\cite{Wegner76}.

The above scaling arguments have used only what in the literature is called
a naive or engineering (with apologies to engineers) dimensional
analysis. For instance, we have assumed that since the conductivity is measured
in units of centimeters to the power $2-d$, it will scale like $\xi^{2-d}$.
In general this type of argument is not correct. The conductivity, for
example, could acquire an `anomalous dimension' $\theta$ by scaling like
$\sigma \sim \xi^{2-d+\theta}\,a^{-\theta}$, where the `wrong' dependence on
the length scale $\xi$ is compensated for by a dependence on some microscopic
length scale $a$. However, for some quantitites, such as $f$
and $\sigma$ above, there are additional arguments that show that they
have zero anomalous dimension so that the above argument is correct.
There is, however, yet another way for scaling relations like
Eq.\ (\ref{eq:2.4b}) to break down, which we will discuss at the end of
the current Section.

Let us now formalize the scaling approach a bit. Suppose we are interested
in an observable $Q$ as a function of $t$, $T$, frequency $\Omega$,
wavenumber $k$, and possibly an external field $h$. The power law
singularities that characterize continuous phase transitions imply
that the singular or scaling part $Q_s$ of the observable is a
generalized homogeneous
function of the appropriate variables. That is, the ususal scaling ansatz,
which typically can be derived by using renormalization group methods, is
\begin{equation}
Q_s(t,T,h;k,\Omega) = b^{-[Q]}\,Q_s(b^{1/\nu}t,b^zT,b^{y_h}h;bk,b^z\Omega)\quad.
\label{eq:2.5}
\end{equation}
Here $b$ is an arbitrary length rescaling factor, and the
number $[Q]$ is called the 
scale dimension of the quantity $Q$. The remaining exponents in 
Eq.\ (\ref{eq:2.5}) have been chosen so that $b$ really is the length 
rescaling factor. For example, if $Q = \xi$ is
the correlation length, then at $T=h=k=\Omega=0$, and using 
$[\xi] = -1$ (as one should if $b$ is a length rescaling factor),
we have $\xi (t)=b\,\xi (b^{1/\nu }t)$. Choosing $b=1/t^\nu$ gives 
$\xi(t) = \xi(1)/t^\nu$, i.e., $\xi$ correctly diverges as $t^{-\nu}$ as 
$t\rightarrow 0$, and $\xi(1)$ is the value of the
correlation length far away from the critical point.

To illustrate the usefulness of these homogeneity laws, let us consider
the conductivity $\sigma$ again, as a function of $t$, $T$, and $\Omega$,
\begin{mathletters}
\label{eqs:2.6}
\begin{equation}
\sigma (t,T,\Omega) = b^{-(d-2)}\,\sigma (b^{1/\nu }t,b^zT,b^z\Omega)\quad.
\label{eq:2.6a}
\end{equation}
By choosing $b=1/t^\nu $, Eq.\ (\ref{eq:2.6a}) gives,
\begin{eqnarray}
\sigma (t,T,\Omega) &=&t^{\nu (d-2)}\,\sigma (1,T/t^{\nu z},\Omega/t^{\nu z})
\nonumber\\
&\equiv &t^{\nu (d-2)}\,F_\sigma (T/t^{\nu z},\Omega/t^{\nu z})\quad,
\label{eq:2.6b}
\end{eqnarray}
and by choosing $b = 1/T^{1/z}$ we obtain,
\begin{eqnarray}
\sigma(t,T,\Omega) &=& T^{(d-2)/z}\,\sigma(t/T^{1/\nu z},1,\Omega/T)
\nonumber\\
&\equiv&T^{(d-2)/z}\,G_{\sigma}(t/T^{1/\nu z},\Omega/T)\quad.
\label{eq:2.6c}
\end{eqnarray}
\end{mathletters}%
Here we have defined the functions $F_{\sigma}(x,y) = \sigma(1,x,y)$ and
$G_{\sigma}(x,y) = \sigma(x,1,y)$. The important point is that, say, at
$\Omega = 0$ the quantity
$\sigma /t^{\nu (d-2)}$ is a function of $T/t^{\nu z}$
alone, rather then of $t$ and $T$ separately. The resulting collapse of
data on a single scaling curve has been historically important for the
confirmation of scaling at thermal phase transitions\cite{Sengers}, and
it is equally useful at QPTs. We will consider an example in
Sec.\ \ref{subsubsec:VI.B.3} below.

Equations\ (\ref{eqs:2.6}) reflect the fact that $t$ is not the only relevant
distance from the critical point. At zero frequency and wavenumber there are 
two relevant variables: $t$ and $T$. For either $t\neq 0$ or
$T\neq 0$ the system is away from the critical surface, and the exponents $\nu$
and $z$ are measures of the relevance of $t$ and $T$, respectively. 
Of course, in any system there are numerous other variables that are 
irrelevant. Let us denote a generic irrelevant variable by $u$. In a
homogeneity law such as Eqs.\ (\ref{eqs:2.6}), these 
irrelevant variables also occur, but they have
negative scale dimensions, $[u]<0$. 
We generalize Eq.\ (\ref{eq:2.6a}) at $\Omega=0$ to
\begin{mathletters}
\label{eqs:2.7}
\begin{equation}
\sigma (t,T,u)=b^{-(d-2)}\sigma (b^{1/\nu }t,b^zT,b^{[u]}u)\quad.
\label{eq:2.7a}
\end{equation}
Choosing $b=1/t^\nu $ now gives
\begin{equation}
\sigma (t,T,u) = t^{\nu (d-2)}\,\sigma (1,T/t^{\nu z},t^{-\nu [u]}u)
                                                                         \quad,
\label{eq:2.7b}
\end{equation}
which for $t\rightarrow 0$ yields (remember $[u]<0$)
\begin{eqnarray}
\sigma (t\rightarrow 0,T,u)&=&t^{\nu (d-2)}\,\sigma (1,T/t^{\nu z},0) 
\nonumber \\
&\equiv &t^{\nu (d-2)}\,F_\sigma (T/t^{\nu z})\quad.
\label{eq:2.7c}
\end{eqnarray}
\end{mathletters}%
We see that close to the critical point, $t\rightarrow 0$, the
dependence on the irrelevant variable drops out. This is a very important
result and is one of the reasons for the universal behavior near continuous
phase transitions.

We conclude this subsection by introducing the concept of a dangerous
irrelevant variable\cite{Fisher}. Let us use Eq.\ (\ref{eq:2.7c}) as an
example. Implicit in the last equality in that equation is the assumption 
that $\sigma (1,T/t^{\nu z},u=0)$ is neither zero nor infinity. This is a
nontrivial assumption which can break down. If it does, then the irrelevant
variable $u$ is called a dangerous irrelevant variable
and actually modifies the scaling behavior, and
scaling equalities, even though it is irrelevant in the technical sense of
the renormalization group. Whether or not a given irrelevant variable is
dangerous with respect to a certain observable can usually not be determined
on general grounds, but requires explicit calculations. In the context of
thermal phase transitions, it is dangerous irrelevant variables
that lead to a breakdown of
hyperscaling above the upper critical dimensionality\cite{PhaseTransitions}. 
In the following sections we will repeatedly see that dangerous irrelevant
variables
play an even more important role in QPTs, mostly due to the small value
of the upper critical dimensionality of the latter. In particular, we will
see in Sec.\ \ref{subsubsec:VI.B.2} that Wegner's scaling law, 
Eqs.\ (\ref{eq:2.4b}), (\ref{eq:2.7c}), is violated at certain 
metal--insulator transitions due to a dangerous irrelevant variable.

\section{Fermionic Field Theory}
\label{sec:III}

The many--fermion problem has a long history, due to its importance with
respect to electrons in condensed matter systems. In order to describe
nonperturbative effects in these systems, such as phase transitions involving
spontaneous symmetry breaking, it is convenient to go to a field theoretic
description. In such a formalism the important underlying symmetries are
most apparent, and, at least formally, it is relatively easy to integrate
out the degrees of freedom that are irrelevant for
describing the physics one wants to focus on. Here we will use standard
field theoretic techniques\cite{NegeleOrland} to sketch the development
of a theory for disordered interacting electrons, following along the
lines of Ref.\ \onlinecite{fermions}.
 
\subsection{Grassmannian field theory}
\label{subsec:III.A}

Since the description of fermions involves anticommuting
variables, any field theory or functional Feynman path integral description
for electrons must be formulated in terms of anticommuting or Grassmann 
variables\cite{Berezin}. 
For the sake of simplicity we consider a model for a homogeneous
electron fluid subject to a random potential that models the quenched
disorder. For most of the effects we want to study, the inclusion of
microscopic details
like the underlying crystal structure and the resulting band structure 
effects, or the explicit inclusion of phonons, is not necessary.
The partition function can then be written\cite{NegeleOrland}
\begin{equation}
Z=\int D[\bar{\psi},\psi ]\exp [S]\quad.
\label{eq:3.1}
\end{equation}
Here the functional integration is with respect to Grassmann valued
fields, $\bar{\psi}$ and $\psi$, and the action $S$ is given by
\begin{mathletters}
\label{eqs:3.2}
\begin{equation}
S=-\int dx\sum\limits_\sigma \bar{\psi}_\sigma (x)\partial _\tau \psi
_\sigma (x)+S_0+S_{dis}+S_{int}\quad.
\label{eq:3.2a}
\end{equation}
We use a $(d+1)$--vector notation, with $x=({\bf x},\tau)$, and 
$\int dx=\int_{V}d{\bf x}\int_{0}^{\beta}d\tau$. ${\bf x}$ denotes
position, $\tau$ imaginary time, V is the system volume, $\beta =1/T$ is
the inverse temperature, and $\sigma$ is the spin label. $S_0$ describes free
electrons with chemical potential $\mu$,
\begin{equation}
S_0=\int dx\sum\limits_\sigma \bar{\psi}_\sigma (x)\left( \frac{\nabla^2}{2m}
+\mu \right) \psi_{\sigma}(x)\quad,
\label{eq:3.2b}
\end{equation}
with $m$ the fermion mass. We will find it useful to go
to a Fourier representation with wavevectors {\bf k} and fermionic Matsubara
frequencies $\omega_n=2\pi T(n+1/2)$,
\footnote[8]{We will denote fermionic frequencies by $\omega_n$, and bosonic
ones by $\Omega_n$ throughout.}
 and a $(d+1)$--vector notation, 
$k=({\bf k},\omega_n)$. Then $S_0$ reads,
\begin{equation}
S_0 = \sum_{k,\sigma}\bar{\psi}_\sigma (k)\left[i\omega_n-\frac{{\bf k}^2}
{2m}+\mu\right] \psi_{\sigma}(k)\quad.  
\eqnum{3.2b'}
\label{eq:3.2b'}
\end{equation}
Here we have redefined $S_0$ to include the first term on the right--hand
side of Eq.\ (\ref{eq:3.2a}).
$S_{dis}$ describes a static random potential, $u({\bf x})$, that couples
to the fermionic number density,
\begin{equation}
S_{dis} = -\int dx\ u({\bf x})\sum_{\sigma}\bar{\psi}_\sigma (x)\,\psi
_\sigma (x)\quad,
\label{eq:3.2c}
\end{equation}
and $S_{int}$ describes a spin--independent two--particle interaction,
\begin{equation}
S_{int} = -\frac{1}{2}\int dx_{1}dx_{2}\ v({\bf x}_1-{\bf x}_2)\sum_{\sigma_1,
                                                                    \sigma_2}
\bar{\psi}_{\sigma _1}(x_1)\bar{\psi}_{\sigma _2}(x_2)\psi _{\sigma
_2}(x_2)\psi _{\sigma _1}(x_1)\quad.
\label{eq:3.2d}
\end{equation}
\end{mathletters}%
The interaction potential $v({\bf x})$ will be discussed further below.
 
For simplicity, the random potential in Eq.\ (\ref{eq:3.2c}) 
is taken to be delta--correlated with a Gaussian distribution. 
Its second moment is,
\begin{equation}
\left\{ u({\bf x})u({\bf y})\right\} _{dis}=\frac 1{\pi N_F\tau_{\rm el}}\,
   \delta ( {\bf x}-{\bf y})\quad, 
\label{eq:3.3}
\end{equation}
where $\left\{\ldots\right\}_{dis}$ denotes the
disorder average. Here $\tau_{\rm el}$ 
is the elastic--scattering mean--free time, and
$N_F$ denotes the bare single--particle density of states (for both spin
projections). More general
random potential models can be used, but any differences between such
generalizations and Eq.\ (\ref{eq:3.3}) are irrelevant for the
long--wavelength effects we are concerned with in this review.
 
For disordered systems, single--particle momentum excitations are not long
lived, but decay exponentially on a time scale given by the elastic 
mean--free time, $\tau_{\rm el}$. 
The important physics on the longest length and time
scales is controlled by two--particle excitations that are soft either
because they are related to conserved quantities or because of a
mechanism related to Goldstone's theorem.
For a detailed discussion of the soft modes in a disordered,
interacting electronic system we refer the reader to 
Ref.\ \onlinecite{fermions}. The main conclusion is
that of all the modes that appear in $S_{int}$, the dominant soft modes are 
those
that involve fluctuations of either the particle number density $n_n$, or
the spin density ${\bf n}_s$, or density
fluctuations, $n_c$, in the particle--particle or Cooper channel. 
In Fourier space,
these densities are given in terms of fermion fields by
\begin{mathletters}
\label{eqs:3.4}
\begin{equation}
n_n(q) = \sqrt{T/V}\,\sum_{k,\sigma}\bar{\psi}_{\sigma}(k)\,\psi_{\sigma}(k+q)
                                                                      \quad,
\label{eq:3.4a}
\end{equation}
 
\begin{equation}
{\bf n}_s(q) = \sqrt{T/V}\sum_{k,\sigma,\sigma^{\prime}}\bar{\psi%
}_\sigma (k)\,{\vec\sigma}_{\sigma \sigma ^{\prime }}\,\psi_{\sigma^{\prime}}
     (k+q)\quad,
\label{eq:3.4b}
\end{equation}
with ${\vec \sigma} = (\sigma_x,\sigma_y,\sigma_z)$ the Pauli matrices, and
\begin{equation}
n_c(q) = \sqrt{T/V}\,\sum_k \psi_{\uparrow}(k)\psi_{\downarrow}(-k+q)\quad. 
\label{eq:3.4c}
\end{equation}
\end{mathletters}%
This means that in Eq.\ (\ref{eq:3.2d}) we should
project onto these fluctuations. 
\footnote[9]{For short--ranged model interactions, this can be done in a
straightforward way\cite{fermions}. A Coulomb interaction requires a
somewhat more elaborate procedure\cite{Coulombfermions}. However, the
final result is given by Eqs.\ (\ref{eqs:3.6}) in either case.}
The net result is that $S_{int}$ can be
split into three terms, each with a different interaction amplitude,
\begin{equation}
S_{int}=S_{int}^{(s)}+S_{int}^{(t)}+S_{int}^{(c)}\quad, 
\label{eq:3.5}
\end{equation}
with the singlet, triplet, and Cooper channel terms simply describing
interactions between the densities given in Eqs.\ (\ref{eqs:3.4}),
\begin{mathletters}
\label{eqs:3.6}
\begin{equation}
S_{int}^{(s)}=-\frac{\Gamma ^{(s)}}2\sum\limits_qn_n(q)\,n_n(-q)\quad, 
\label{eq:3.6a}
\end{equation}
 
\begin{equation}
S_{int}^{(t)}=\frac{\Gamma ^{(t)}}2\sum\limits_q{\bf n}_s(q)\cdot{\bf n}_s(-q)
                                                                      \quad,
\label{eq:3.6b}
\end{equation}

\begin{equation}
S_{int}^{(c)}=-\frac{\Gamma ^{(c)}}2\sum\limits_qn_c(q)\,n_c(-q)\quad. 
\label{eq:3.6c}
\end{equation}
\end{mathletters}%
Here $\Gamma^{(i)}$, $i=s,t,c,$ are singlet, triplet and Cooper channel
interaction amplitudes, which are in turn given by angular averages of the
interaction potential at specified momenta.
 
Since the system contains quenched disorder, it is necessary to average the
free energy or $\ln Z$. This is accomplished by means of the 
replica trick\cite{ReplicaTrick}, which is based on the formal identity
\begin{equation}
\ln Z=\lim\limits_{n\rightarrow 0}\ (Z^n-1)/n\quad.
\label{eq:3.7}
\end{equation}
Introducing $n$ identical replicas of the system (with $n$ an
integer), labeled by the index $\alpha$, and carrying out the disorder
average, we obtain
\begin{equation}
\tilde{Z}\equiv\{Z^n\}_{dis}=\int D[\bar{\psi}^\alpha ,\psi ^\alpha ]\,
             e^{\tilde{S}}\quad.
\label{eq:3.8}
\end{equation}
We again separate $\tilde{S}$ into free, disordered and
interacting parts,
\begin{mathletters}
\label{eqs:3.9}
\begin{equation}
\tilde{S}=\sum\limits_{\alpha =1}^n\,(\tilde{S}_0^\alpha +\tilde{S}%
_{dis}^\alpha +\tilde{S}_{int}^\alpha )\quad. 
\label{eq:3.9a}
\end{equation}
$\tilde{S}_0^\alpha $ and $\tilde{S}_{int}^\alpha $ are given by
Eqs.\ (\ref{eq:3.2b'}) and (\ref{eq:3.5}) and (\ref{eqs:3.6}), 
respectively with $\psi \rightarrow \psi^\alpha $ and $S_{dis}^\alpha $ 
is given by,
\begin{equation}
\tilde{S}_{dis}^\alpha = \frac 1{2\pi N_F\tau_{\rm el}}\sum\limits_{\beta
=1}^n\sum\limits_{\{{\bf k}\}}\sum_{n,m}\sum\limits_{\sigma ,\sigma ^{\prime
}}\delta _{{\bf k}_1+{\bf k}_3,{\bf k}_2+{\bf k}_4} \\
\ \bar{\psi}_{n\sigma }^\alpha ({\bf k}_1)\psi _{n\sigma }^\alpha ({\bf k%
}_2)\bar{\psi}_{m\sigma ^{\prime }}^\beta ({\bf k}_3)\psi _{m\sigma ^{\prime
}}^\beta ({\bf k}_4)\quad.
\label{eq:3.9b}
\end{equation}
\end{mathletters}%
At the end of all calculations, one takes the limit $n\rightarrow 0$ (assuming
that both the analytic continuation from integer to real $n$ and the limit 
exist) to obtain the desired averages for quenched disorder.

\subsection{Order parameter field theories}
\label{subsec:III.B}

Starting with the pioneering work of Landau\cite{Landau}, it emerged over
the years that an expedient way to describe any
phase transition is to formulate an effective field theory, the LGW
functional, entirely in
term of an appropriate order parameter, and to bury the information about
all other degrees of freedom in the vertices of that order parameter
field theory. As we will see, this approach is more problematic for
quantum phase transitions than for thermal ones. The reason is that
the LGW approach hinges in part on an implicit assumption, namely that
the order parameter fluctuations are the only soft modes in the system.
Although this is not true for most of the quantum phase transitions we will be
interested in, it turns out that the LGW approach nevertheless provides
an economical way of solving the problem, provided one adequately deals
with the ensuing complications. In the current subsection we therefore
show how to derive LGW functionals of ferromagnetic, antiferromagnetic,
and superconducting order parameters, respectively; for the time being
without worrying about any additional soft modes. The question of an
order parameter description of metal--insulator transitions is more
intricate, and does require a thorough examination of the symmetry 
and soft mode properties
of the underlying fermionic field theory, Sec.\ \ref{subsec:III.A}, as a
prerequisite. This analysis will be given in Sec.\ \ref{subsec:III.C}
below.

\subsubsection{Magnetic order parameters}
\label{subsubsec:III.B.1}

It is well known that the model defined in Sec.\ \ref{subsec:III.A}
contains a ferromagnetic phase. Large values of the coupling constant
$\Gamma^{(t)}$ and small values of $T$ favor ferromagnetism, while
small values of $\Gamma^{(t)}$ and large values of $T$ favor paramagnetism.
The appropriate order parameter for the transition
between these two phases is the magnetization ${\bf m}$ or the expectation
value of the spin density defined in Eq.\ (\ref{eq:3.4b}):
${\bf m} = \langle {\bf n}_s({\bf x})\rangle$. Technically, the derivation of
a LGW functional is easily achieved by means of a trick due to
Stratonovich and Hubbard\cite{HubbardStratonovich}, 
which is based on the identity
\begin{equation}
\int D[{\bf M}]\ e^{-(\Gamma^{(t)}/2)\int dx\ {\bf M}(x)\cdot
   {\bf M}(x) + \Gamma^{(t)}\int dx\ {\bf M}(x)\cdot
      {\bf n}_s(x)} = {\rm const}\times e^{S_{int}^{(t)}}\quad.
\label{eq:3.10}
\end{equation}
Here ${\bf M}(x)$ is a classical (i.e., complex number valued) auxiliary
field that couples linearly to the spin density. The strategy is now to
apply this Gaussian decoupling procedure, which is referred to as a
Hubbard--Stratonovich decoupling or as `uncompleting the square', to
$S_{int}^{(t)}$ while leaving all other terms in the action unchanged.
We thus write the action as
\begin{equation}
S = S_{ref} + \ln\int D[{\bf M}]\ e^{-(\Gamma^{(t)}/2)\int dx\ {\bf M}(x)\cdot
          {\bf M}(x) + \Gamma^{(t)}\int dx\ {\bf M}(x)\cdot{\bf n}_s(x)}\quad,
\label{eq:3.11}
\end{equation}
where the $S_{ref}$ contains all pieces of the action other than 
$S_{int}^{(t)}$. We will refer to the system described by $S_{ref}$ as the
`reference ensemble'.

By applying the Hubbard--Stratonovich transformation explained above,
and formally performing the fermionic integral,
the replicated partition function, Eq.\ (\ref{eq:3.8}), can be written as
\begin{mathletters}
\label{eqs:3.12}
\begin{equation}
\tilde{Z} = e^{-nF_0/T}\int\prod_{\alpha}\,D[{\bf M}^{\alpha}]
                      \exp \bigl(-\Phi [{\bf M}]\bigr)\quad,
\label{eq:3.12a}
\end{equation}
with $F_0$ the noncritical part of the free energy. 
\footnote[10]{$F_0$ is noncritical as long as the reference ensemble does not
undergo a phase transition of its own. This is not an entirely trivial point,
as perturbation theory within the reference ensemble will in general create
a nonvanishing spin--triplet interaction, even though there was none in the
bare system. For all of our order parameter theories we assume that the
reference ensemble is far from any phase transitions, since else the mode
separation that is implicit in Eq.\ (\ref{eq:3.11}) breaks down, and nothing
is gained by the Hubbard--Stratonovich transformation.}
The LGW
functional $\Phi$ reads
\footnote[11]{We denote the microscopic fermionic action by $S$, LGW functionals
that depend only on the classical order parameter field by $\Phi$, and
effective fermion actions by $\cal A$, respectively.}
\begin{eqnarray}
\Phi [{\bf M}] &=&\frac{\Gamma _t^{(t)}}2\sum\limits_\alpha \int dx\ {\bf M}
^\alpha (x)\cdot {\bf M}^\alpha (x)  
\nonumber \\
&&-\ln \left\langle \exp \left(-\Gamma ^{(t)}\sum_{\alpha}\int dx\ {\bf M}
^\alpha (x)\cdot {\bf n}_s^\alpha (x)\right)\right\rangle_{ref}\quad.
\label{eq:3.12b}
\end{eqnarray}
Here $\langle\ldots\rangle_{ref}$ denotes an average taken with respect to
the reference ensemble. A useful interpretation of the second contribution
to the LGW functional is obtained by realizing that it represents the
free energy of the reference ensemble in the presence of an external
magnetic field that is proportional to the order parameter field ${\bf M}$.
Also notice that the average of ${\bf M}$ is proportional to the
magnetization ${\bf m}$, by virtue of the linear coupling between ${\bf M}$
and ${\bf n}_s$.

For clean systems, and a more restricted reference ensemble, the above 
derivation of an LGW functional for quantum
ferromagnets is due to Hertz\cite{Hertz}. Our derivation\cite{fmdirty} has
generalized his approach to include disorder, and electron interaction
in multiple channels.
The standard way to proceed with a discussion of the 
paramagnet--to--ferromagnet transition is a Landau expansion or 
expansion of the LGW functional
$\Phi$ in powers of the order parameter field ${\bf M}$. 
The structure of this expansion is
\begin{eqnarray}
\Phi[{\bf M}]&=&\sum_{l=2}^{\infty} \Phi_l[{\bf M}]
\nonumber\\
&=&{1\over 2}\sum_{\alpha} \int dx_1\,dx_2\ X^{(2)}_{ab}(x_1,x_2)\,
        M_a^{\alpha}(x_1)\,M_b^{\alpha}(x_2)
\nonumber\\
 &&+\ {1\over 3!}\sum_{\alpha} \int dx_1\,dx_2\,dx_3\ X^{(3)}_{abc}(x_1,x_2,x_3)
                \,M_a^{\alpha}(x_1)\,M_b^{\alpha}(x_2)\,M_c^{\alpha}(x_3)
\nonumber\\
  &&-\ {1\over 4!}\sum_{\alpha,\beta} \int dx_1\,dx_2\,dx_3\,dx_4\
               X^{(4)\,\alpha\beta}_{abcd}(x_1,x_2,x_3,x_4)
 \, M_a^{\alpha}(x_1)\,M_b^{\alpha}(x_2)\,M_c^{\beta}(x_3)\,M_d^{\beta}(x_4)
  + O(M^5)\quad,
\label{eq:3.12c}
\end{eqnarray}
\end{mathletters}%
The vertex functions $X^{(l)}$ are given in terms of the spin density
correlations of the reference ensemble, and can be calculated within
any theory of interacting disordered electrons. This we will come back to
in Sec.\ \ref{sec:IV} below, where we will see that
this Landau expansion is ill--behaved in the sense that the vertex functions
cannot be localized in space and time.

For an antiferromagnet the interesting fluctuations are those of the
staggered magnetization. Our simple continuum model, Sec.\ \ref{subsec:III.A},
actually has no antiferromagnetic phase. However, order parameter
theories for antiferromagnets can be derived from more complicated microscopic
theories\cite{Anderson}. The net result is again a LGW functional of the
form of Eq.\ (\ref{eq:3.12c}), with the fluctuating magnetization ${\bf M}$
replaced by a different vector order parameter ${\bf N}$, whose average is
proportional to the staggered magnetization. The structure of
the vertex functions in this case is different from the ferromagnetic one.
We will model these vertex functions in Sec.\ \ref{subsec:IV.B}.

\subsubsection{Superconducting order parameter}
\label{subsubsec:III.B.3}

For superconductors, the procedure is completely analogous to that in the
ferromagnetic case, except that one uses a Hubbard--Stratonovich decoupling
on the Cooper channel interaction, $S_{int}^{(c)}$. We thus write
\begin{equation}
{\tilde S} = {\tilde S}_{ref} + {\tilde S}_{int}^{(c)}\quad,
\label{eq:3.13}
\end{equation}
with ${\tilde S}_{int}^{(c)}$ the replicated generalization of 
Eq.\ (\ref{eq:3.6c}), and proceed as in
Sec.\ \ref{subsubsec:III.B.1}, with the only difference being that we now
have a complex valued order paramter that we denote by $\Psi$. The LGW
functional reads
\begin{equation}
\Phi[\Psi] = -\Gamma^{(c)}\sum_{\alpha}\int dq\ \vert\Psi^{\alpha}(q)\vert^2 
      - \ln\left\langle e^{-\Gamma^{(c)}\sum_{\alpha} 
            \int dq\,\bigl({\Psi^{\alpha}}^*(q)\,n_c^{\alpha}(q) 
                       + {\Psi}^{\alpha}(q)\,{\bar n}_c^{\alpha}(q)\bigr)}
     \right\rangle_{ref}\quad.
\label{eq:3.14}
\end{equation}
The Landau expansion in powers of the order parameter is again analogous
to the ferromagnetic case, Eq.\ (\ref{eq:3.12c}), except that due to
gauge invariance only even powers of $\Psi$ appear, and due to the
complex valuedness of $\Psi$ the vertex functions are scalars rather
than tensors.
This order parameter field theory will be discussed in Sec.\ \ref{sec:V}
below.

\subsection{The nonlinear sigma--model}
\label{subsec:III.C}

Before we continue the discussion of the effective field theories derived
in the previous subsection, it is useful to discuss the properties of the
underlying fermionic field theory in some detail, in particular with respect
to its symmetry properties and soft mode structure. This is very helpful
for an understanding of the properties of the order parameter theories,
and it is necessary in order to derive an order parameter theory for
metal--insulator transitions. The symmetry analysis of the fermionic
theory is very technical and involved, and the details can be found in 
Ref.\ \onlinecite{fermions}. Here our goal is to explain the general structure
of the theory, and the logic of the analysis, by exploiting a remarkable
analogy between classical spin models for Heisenberg ferromagnets and
disordered electrons that was first noted and used by 
Wegner\cite{Wegner79}. 

\subsubsection{Digression: The nonlinear sigma--model for classical 
                                              Heisenberg ferromagnets}
\label{subsubsec:III.C.1}

Let us consider, as a model for a classical Heisenberg ferromagnet,
\footnote[12]{The fact that classical ferromagnets are useful here
for pedagogical purposes, while in Secs.\ \ref{subsubsec:III.B.1} and
\ref{subsec:IV.A} we discuss quantum ferromagnets, is accidental.} 
an $O(N)$ symmetric $\phi^4$--theory with a magnetic field $h$ in the 
$1$--direction. The action
\begin{mathletters}
\label{eqs:3.15}
\begin{equation}
S[{\vec\phi},h] = \int d{\bf x}\,\left[r\bigl({\vec\phi}({\bf x})\bigr)^2
               + c\bigl(\nabla{\vec\phi}({\bf x})\bigr)^2\right]
 + u \int d{\bf x}\,\left({\vec\phi}({\bf x})\cdot{\vec\phi}({\bf x})\right)^2
  - h\int d{\bf x}\,\phi_1({\bf x}) \quad,
\label{eq:3.15a}
\end{equation}
determines the partition function
\begin{equation}
Z[h] = \int D[{\vec\phi}\,]\,e^{-S[{\vec\phi},h]}\quad,
\label{eq:3.15b}
\end{equation}
\end{mathletters}%
and $r$, $c$, and $u$ are real--valued coefficients that span the parameter
space of the theory. The following discussion of this action, along standard 
lines\cite{ZJ}, will turn out to be very useful as an analogy for the 
fermionic theory we are interested in.

The crucial property of the action, Eq.\ (\ref{eq:3.15a}), that we want to
exploit is its symmetry: For zero external magnetic field, $h=0$, the
action is invariant under $O(N)$ rotations of the field ${\vec\phi}$. Let us
consider a particular rotation, between $\phi_1$ and $\phi_i$
with $2\leq i\leq N$, through an infinitesimal angle $\theta$,
\begin{eqnarray}
\phi_1 \rightarrow \phi_1^{\prime} = \phi_1 + \delta\phi_1 
                                   = \phi_1 + \theta\,\phi_i\quad,
\nonumber\\
\phi_i \rightarrow \phi_i^{\prime} = \phi_i + \delta\phi_i 
                                   = \phi_i - \theta\,\phi_1\quad.
\label{eq:3.16}
\end{eqnarray}
We also introduce a source ${\vec J}({\bf x})$ for the 
${\vec\phi}$--field,
\footnote[13]{This source can be thought of as a spatial modulation of the
magnetic field $h$.} 
and consider the generating functional
\begin{equation}
Z[h,{\vec J}\,] = \int D[{\vec\phi}\,]\,e^{-S[{\vec\phi},h] + \int d{\bf x}\,
                {\vec J}({\bf x})\cdot{\vec\phi}({\bf x})}\quad.
\label{eq:3.17}
\end{equation}
If we change integration variables from ${\vec\phi}$ to ${\vec\phi}^{\prime}$,
then the Jacobian of that transformation is equal to one. To linear
order in the infinitesimal group parameter $\theta$ we thus obtain
\begin{equation}
\int D[{\vec\phi}\,]\,\left(-\delta S[{\vec\phi},h] + \int d{\bf x}\,{\vec J}
       ({\bf x})\cdot\delta{\vec\phi}({\bf x})\right)\,e^{-S[{\vec\phi},h]}
                                                             = 0 \quad,
\label{eq:3.18}
\end{equation}
where $\delta S = S[{\vec\phi}^{\prime}] - S[{\vec\phi}\,]$.
By differentiating with respect to $J_i({\bf x})$ and setting ${\vec J}=0$
after the differentiation we obtain $N-1$ Ward identities,
\begin{equation}
\langle \delta S\cdot\phi_i({\bf x})\rangle + \langle\delta\phi_i({\bf x})
             \rangle = 0\quad,\quad (i=2,\ldots,N)\quad,
\label{eq:3.19}
\end{equation}
where the brackets denote an average with respect to the action $S$.
The only part of $S$ that is not invariant under the rotation
is the magnetic field term. The Ward identities, Eq. (\ref{eq:3.19}), thus 
take the form of a relation between the transverse (with respect to the
magnetic field) two--point correlation functions and the longitudinal
one--point function,
\begin{mathletters}
\label{eqs:3.20}
\begin{equation}
h \int d{\bf y}\ \langle\phi_i({\bf x})\,\phi_i({\bf y})\rangle
             = \langle\phi_1({\bf x})\rangle\quad.
\label{eq:3.20a}
\end{equation}
Since the left--hand side of this identity is $h$ times the homogeneous
transverse susceptibility $\chi_t$, while the right--hand side is equal
to the magnetization $m$, this can be rewritten as
\begin{equation}
\chi_t = m/h\quad.
\label{eq:3.20b}
\end{equation}
\end{mathletters}%
We thus obtain the well--known result that the transverse zero--field
susceptibility diverges everywhere in the ordered phase, i.e. whenever $m>0$. 
This is a particular example of Goldstone's theorem: Since there is an
$(N-1)$--parameter
continuous symmetry ($O(N)$ in our case) that is spontaneously broken
($m=\langle\phi_1\rangle_{h\rightarrow 0}\neq 0$ in the ordered phase),
there are $N-1$ massless excitations (the transverse $\phi$--$\phi$
correlation functions in the limit $h\rightarrow 0$).

In ordinary perturbation theory the masslessness of the $N-1$ Goldstone
modes is not manifest, but rather has to be established by means of
explicit calculations order by order\cite{Ma}. 
This is clearly undesirable, and raises the question of whether one can
reformulate the theory so that this qualitative feature is explicitly
displayed. This can indeed be done, and the result takes the form of
a so--called nonlinear sigma--model\cite{ONSigmaModel}. The basic idea
is the realization that the massless Goldstone modes correspond to
purely transverse fluctuations of the vector fields, as the above
derivation of the Ward identity demonstrates.
\footnote[14]{Notice that the
$O(N)$ transformation leaves the modulus of the vector field invariant.}
Let us therefore decompose the vector field $\vec\phi$ into its modulus 
$\rho$ and a unit vector field $\hat\phi$,
\begin{equation}
{\vec\phi}({\bf x}) = \rho({\bf x})\,{\hat\phi}({\bf x})\quad,\quad
                        {\hat\phi}^2({\bf x})\equiv 1\quad,
\label{eq:3.21}
\end{equation}
Let ${\hat\psi}$ be a particular fixed unit $N$--vector. Then any $\hat\phi$
can be generated from ${\hat\psi}$ by means of an $O(N)$--rotation. The
set of the $\hat\phi$ is thus isomorphic to $O(N)$ modulo the subgroup
that leaves ${\hat\psi}$ invariant, which is $O(N-1)$. The $\hat\phi$
therefore provide a representation of the homogeneous space $O(N)/O(N-1)$.
In terms of $\rho$ and $\hat\phi$ the action reads,
\begin{mathletters}
\label{eqs:3.22}
\begin{equation}
S[\rho,\hat\phi] = \int d{\bf x}\,\biggl[c\rho^2({\bf x})
      \left(\nabla{\hat\phi}({\bf x})
            \right)^2 + c\left(\nabla\rho({\bf x})\right)^2
 + r\rho^2({\bf x})\biggr] + u\int d{\bf x}\,\rho^4({\bf x})
  - h\int d{\bf x}\,\rho({\bf x}){\hat\phi}_1({\bf x})\quad,
\label{eq:3.22a}
\end{equation}
and switching from the functional integration variables $\vec\phi$ to
$(\rho,\hat\phi)$ leads to a Jacobian or invariant measure
\begin{equation}
I[\rho] = \prod_{\bf x} \rho^{N-1}({\bf x})\quad.
\label{eq:3.22b}
\end{equation}
\end{mathletters}%
The important point is that in this formulation of the
theory, the field $\hat\phi$ appears only in
conjunction with two gradient operators. $\hat\phi$ represents the $N-1$ soft
Goldstone modes of the problem, while $\rho$ represents the massive modes.
Now we parametrize $\hat\phi$,
\begin{mathletters}
\label{eqs:3.23}
\begin{equation}
{\hat\phi}({\bf x}) = \bigl(\sigma ({\bf x}),{\vec\pi}({\bf x})
                          \bigr)\quad,
\label{eq:3.23a}
\end{equation}
where
\begin{equation}
\sigma ({\bf x}) = \sqrt{1-{\vec\pi}^2({\bf x})}\quad.
\label{eq:3.23b}
\end{equation}
\end{mathletters}%
We split off the expectation value of the massive $\rho$--field,
$\rho({\bf x}) = R + \Delta\rho({\bf x})$, with
$R = \langle\rho({\bf x})\rangle$, and expand in powers of $\vec\pi$ and
$\Delta\rho$. Rescaling the coupling constants with appropriate powers of
$R$, the action can be written
\begin{mathletters}
\label{eqs:3.24}
\begin{equation}
S[\rho,{\vec\pi}] = S_{NL\sigma M}[{\vec\pi}] + \Delta S[\rho,{\vec\pi}]\quad.
\label{eq:3.24a}
\end{equation}
Here
\begin{equation}
S_{NL\sigma M}[{\vec\pi}] = \frac{1}{t}\int d{\bf x}\,\left[
             \left(\nabla{\vec\pi}({\bf x})\right)^2
                            + \left(\nabla\sigma({\bf x})\right)^2\right]
- h\int d{\bf x}\,\sigma{(\bf x})\quad,
\label{eq:3.24b}
\end{equation}
is the action of the $O(N)/O(N-1)$ nonlinear sigma--model, 
which derives its name
from the notation used in Eqs.\ (\ref{eqs:3.23}), and
\begin{eqnarray}
\Delta S[\rho,{\vec\pi}] = r \int d{\bf x}\,\left(\Delta\rho ({\bf x})\right)^2
            + c \int d{\bf x}\,\left(\nabla \Delta\rho ({\bf x})\right)^2
  + 4Ru \int d{\bf x}\,\left(\Delta\rho ({\bf x})\right)^3
    &+&u \int d{\bf x}\,\left(\Delta\rho ({\bf x})\right)^4
\nonumber\\
 &+&O\left(\Delta\rho\,\sigma,\Delta\rho\left(\nabla{\hat\phi}
                                                \right)^2\right)\quad,
\label{eq:3.24c}
\end{eqnarray}
\end{mathletters}%
contains the corrections to it, which are all massive.
If we neglect all fluctuations of the massive $\rho$--field, 
then we are left with
the $O(N)/O(N-1)$ nonlinear sigma--model in the usual parametrization.

The separation of soft and massive modes is now complete, and will be
preserved order by order in perturbation theory. We will
explain in the next subsection how to perform an analogous separation of modes
for the fermionic action of Sec.\ \ref{subsec:III.A}. Before we do that,
however, let us briefly discuss a way to rewrite Eq.\ (\ref{eq:3.24b}) 
that will be very useful later. We enforce the constraint ${\hat\phi}^2 = 1$
explicitly by means of a Langrange multiplier field $\lambda$,
\begin{mathletters}
\label{eqs:3.25}
\begin{equation}
S_{NL\sigma M}[{\hat\phi},\lambda] = \frac{1}{t}\int d{\bf x}\,\left[
 \left(\nabla{\hat\phi}({\bf x})\right)^2 + \lambda({\bf x})\,\left(
    {\hat\phi}^2({\bf x}) - 1\right)\right]\quad.
\label{eq:3.25a}
\end{equation}
Usually one integrates out $\lambda$, which eliminates $\sigma$ in terms
of ${\vec\pi}$, and expands in powers of ${\vec\pi}$. This leads to a
$(2+\epsilon)$--expansion for the Heisenberg transition. Alternatively,
however, one can integrate out the ${\vec\pi}$ fields to obtain an
effective action ${\cal A}_{eff}$ in terms of $\sigma$ and $\lambda$,
\begin{equation}
{\cal A}_{eff}[\sigma,\lambda] = \frac{1}{t}\int d{\bf x}\,\left[
  \left(\nabla\sigma({\bf x})\right)^2 + \left(\sigma^2({\bf x}) - 1\right)\,
  \lambda({\bf x})\right] + \frac{1}{2}\,(N-1)\,\Tr\ln\left(-\nabla^2 + \lambda
             ({\bf x})\right)\quad.
\label{eq:3.25b}
\end{equation}
\end{mathletters}%
Now look for a homogeneous saddle--point solution. Denoting the saddle--point
values of the fields again by $\sigma$ and $\lambda$, respectively, we find
\begin{eqnarray}
\lambda\sigma &=& 0\quad,
\nonumber\\
\sigma^2 &=& 1 - (N-1)\,t\,\frac{1}{V}
               \sum_{\bf p} \frac{1}{{\bf p}^2 + \lambda}\quad.
\label{eq:3.26}
\end{eqnarray}
These saddle--point equations correctly reproduce the exact critical
behavior for $d>4$, which is mean--field like. They also yield the exact
critical behavior for all $d>2$ in the limit $N\rightarrow\infty$ (with
$tN$ held fixed)\cite{ZJ}.

\subsubsection{Symmetry properties of the fermion model}
\label{subsubsec:III.C.2}

It is possible to rewrite the fermionic action of Sec.\ \ref{subsec:III.A}
in complete analogy to the treatment of the much simpler $\phi^4$--theory 
in the previous subsection. The first question that arises in this
context is which symmetry group takes the place of $O(N)$. For this
purpose, it is convenient to rewrite the fermionic theory in terms of
four--component bispinors $\eta$ and $\eta^+$ that are defined 
by\cite{EfetovLarkinKhmelnitskii}
\begin{mathletters}
\label{eqs:3.27}
\begin{equation}
\eta_{n}^{\alpha}({\bf x}) = 
     \frac{1}{\sqrt{2}}\left(
    \begin{array}{c}\bar{\psi}_{n\uparrow}^{\alpha}({\bf x}) \\
                    \bar{\psi}_{n\downarrow}^{\alpha}({\bf x})\\
                    \psi_{n\downarrow}^{\alpha}({\bf x})\\
                    -\psi_{n\uparrow}^{\alpha}({\bf x})\end{array}
                                         \right)\quad,
\label{eq:3.27a}
\end{equation}
and
\begin{equation}
(\eta^+)_n^\alpha ({\bf x}) = i(C\eta)_n^\alpha ({\bf x})
     = \frac{i}{\sqrt{2}}\left(
    \begin{array}{c}-{\psi}_{n\uparrow}^{\alpha}({\bf x}) \\
                    -{\psi}_{n\downarrow}^{\alpha}({\bf x})\\
                    \bar{\psi}_{n\downarrow}^{\alpha}({\bf x})\\
                    -\bar{\psi}_{n\uparrow}^{\alpha}({\bf x})\end{array}
                                         \right)\quad,
\label{eq:3.27b}
\end{equation}
where
\begin{equation}
^{ij}C_{nm}^{\alpha\beta} = \delta_{nm}\,\delta_{\alpha\beta}\,c_{ij}\quad,
\label{eq:3.27c}
\end{equation}
with $c$ the charge--conjugation matrix
\begin{equation}
c = \left(\begin{array}{cc}0&s_2\\
                           s_2&0\end{array}\right)
  = i\tau_1\otimes s_2\quad.
\label{eq:3.27d}
\end{equation}
\end{mathletters}%
The four degrees of freedom represented by the bispinor are the
particle--hole or number density degrees of freedom, and the two spin
degrees of freedom. We have also defined a basis in spin--quaternion space
as $\tau_r\otimes s_i$ $(r,i=0,1,2,3)$, with $\tau_0=s_0$ the $2\times 2$
identity matrix, and $\tau_j=-s_j=-i\sigma_j$ $(j=1,2,3)$, with $\sigma_j$ the
Pauli matrices. In this basis, the channels $r=0,3$ and $r=1,2$ describe the
particle--hole and particle--particle degrees of freedom, and the
channels $i=0$ and $i=1,2,3$ describe the spin--singlet and spin--triplet,
respectively. In the space of bispinors we further define a scalar
product,
\begin{equation}
\sum_n \sum_{\alpha} \tr\,\Bigl( (\eta^+)_n^{\alpha} \otimes \eta_n^{\alpha}
       \Bigr)\equiv \left(\eta,\eta\right)\quad.
\label{eq:3.28}
\end{equation}

Now let $\eta$ be transformed by means of an operator $\hat T$:
$\eta \rightarrow {\hat T}\eta$.
By rewriting the action in terms of the bispinors, it is easy to see
that the free fermion action $S_0$, Eq.\ (\ref{eq:3.2b'}),
is invariant under any transformations of the bispinors that leave the
metric $(\eta,\eta)$ invariant, except for the part that is proportional
to the external frequency, $i\omega$. The
disorder part of the action $S_{dis}$, Eq.\ (\ref{eq:3.9b}), is also
invariant under such transformations. Remembering that $\eta^+$ is related to
$\eta$ by means of the charge conjugation matrix $C$, Eq.\ (\ref{eq:3.27c}),
and using $C^{\,T} = C^{-1}$, we find that in order
to leave $(\eta,\eta)$ invariant, ${\hat T}$ must obey
\begin{equation}
{\hat T}^{\,T}\,C\,{\hat T} = C\quad.
\label{eq:3.29}
\end{equation}
For a system with $2N$ frequency labels ($N$ positive ones, including $0$, and
$N$ negative ones), and $n$ replicas, Eq.\ (\ref{eq:3.29}) defines a
representation of the
symplectic group ${\rm Sp}(8Nn,{\cal C})$ over the complex
numbers $\cal C$\cite{Wybourne}. ${\rm Sp}(8Nn,{\cal C})$ thus plays
a role analogous to that of $O(N)$ in $\phi^4$--theory, and the external
frequency in $S_0$ can be thought of as being analogous to the external 
magnetic field in the latter.
\footnote[15]{This is true even though $i\omega$ couples to a term that is
bilinear in $\psi$ or $\eta$, while $h$ in Sec.\ \ref{subsubsec:III.C.1}
couples linearly to $\phi$.}
The electron--electron interaction, which also breaks the symmetry,
turns out to be very similar to $i\omega$ in its effect, see below.

The next question is whether there is a phase where the symplectic 
symmetry at zero frequency
is spontaneously broken, and if so, what plays the role of the order
parameter and the Goldstone modes. To see this it is convenient to
define a Grassmannian matrix field $B_{12}$, with
$1\equiv (n_1,\alpha_1)$, etc, by
\begin{mathletters}
\label{eqs:3.30}
\begin{equation}
B_{12}({\bf x}) = \eta^{+}_{1}({\bf x})\otimes\eta_{2}({\bf x})\quad,
\label{eq:3.30a}
\end{equation}
or, in Fourier space and with all components written out,
\begin{equation}
{_i{^jB}}_{nm}^{\alpha\beta}({\bf q}) = \sum_{{\bf k}}\,{_i\eta}_{n}^{\alpha}
({\bf k})\ {^j\eta}_{m}^{\beta}({\bf k}+{\bf q})\quad,
\label{eq:3.30b}
\end{equation}
where $^i\eta$ denotes the elements of $\eta$, and $_i\eta$ those of
$\eta^+$. It is often convenient to expand the $B$ into the spin--quaternion
basis defined above,
\begin{equation}
B_{12} = \sum_{r=0}^{3}\sum_{i=0}^{3} {^i_rB_{12}}\,\left(\tau_r\otimes s_i
             \right)\quad.
\label{eq:3.30c}
\end{equation}
\end{mathletters}%
The set of $B$ is isomorphic to a set of classical, i.e. complex number
valued, matrix fields. One can thus introduce a classical matrix field
$Q$, and constrain $B$ to $Q$ by means of a Langrange multiplier field,
$\Lambda$. The fermionic degrees of freedom can then be integrated out,
and one obtains a theory in terms of the matrix fields $Q$ and $\Lambda$.
In terms of these objects, the density of states, $N$, as a function of
energy or frequency $\omega$ measured from the Fermi surface is
\begin{equation}
N(\epsilon_F + \omega) = \frac{4}{\pi}\,{\rm Re}\,\Bigl\langle{^0_0 Q}_{nn}
        ({\bf x})\Bigr\rangle\Bigr\vert_{i\omega_n\rightarrow\omega + i0}
                                                                \quad.
\label{eq:3.31}
\end{equation}
Similarly, the number density susceptibility
$\chi_n$, and the spin density susceptibility, $\chi_s$, are given by
\begin{equation}
\chi^{(i)}({\bf q},\Omega_n) = 16T\sum_{m_1,m_2}\sum_{r=0,3}\left\langle
{_r^i(\Delta Q)}_{m_1-n,m_1}^{\alpha\alpha}({\bf q})\,
{_r^i(\Delta Q)}_{m_2,m_2+n}^{\alpha\alpha}(-{\bf q})
      \right\rangle\quad ,
\label{eq:3.32}
\end{equation}
with $\Delta Q = Q - \langle Q \rangle$, and $\chi^{(0)} = \chi_n$ and
$\chi^{(1,2,3)} = \chi_s$. $\Omega_n = 2\pi Tn$ is a bosonic Matsubara
frequency.

Under transformations of the bispinors, the $Q$ and $\Lambda$ transform
accordingly. One can then derive Ward identities in analogy to
Eqs.\ (\ref{eq:3.18}) - ({\ref{eq:3.20a}), with the vector field ${\vec\phi}$
replaced by the matrix field $Q$, and the source $J$ also being a matrix.
The detailed calculation shows that the role of the transverse fields
$\phi_i$ ($i=2,\ldots,N$) is played by the matrix elements $Q_{nm}$ with
$nm<0$, i.e. by products of fermion fields whose frequency indices have
different signs, while the analogs of the longitudinal field are matrix
elements with $nm>0$. The 
Goldstone mode equation analogous to Eq.\ (\ref{eq:3.20b}) takes the form,
\begin{equation}
\left\langle{^{0}_{0}Q}_{n_1 n_2}^{\alpha\beta}({\bf k})\,
   {^{0}_{0}Q}_{n_1 n_2}^{\alpha\beta}({\bf -k})\right\rangle
                                                    \bigg\vert_{{\bf k}=0}
 = \frac{\pi N(\epsilon_F)}{16\vert\Omega_{n_1 - n_2}\vert}\qquad,\qquad
        (n_1n_2<0)\quad.
\label{eq:3.33}
\end{equation}
Here $N(\epsilon_F)$ is the exact density of states at the Fermi level.
The salient point is that, as long as $N(\epsilon_F)>0$,
the $Q$-$Q$ correlation function at zero momentum
diverges like $1/\vert\Omega_{n_1 - n_2}\vert$.
We have therefore identified
$^{0}_{0}Q_{n_1 n_2}^{\alpha\alpha}$ for $n_1 n_2 <0$ as a soft mode.
As written, Eq.\ (\ref{eq:3.33}) is valid only for noninteracting
electrons. However, an analysis of the interaction term shows that it
does not spoil the property of the $Q$--$Q$ correlation function on
the left--hand side of Eq.\ (\ref{eq:3.33}) being massless as long as
the density of states at the Fermi level is nonzero. Further, one can
invoke additional symmetries of the action to show that Eq.\ (\ref{eq:3.33})
also holds for $^i_rQ$ with $i,r\neq 0$. The soft modes in the
particle--hole channel ($r=0,3$) and particle--particle channel ($r=1,2$)
are often called `diffusons', and `cooperons', respectively. They are all
soft in the absence of physical processes that break the additional symmetries
which link the general $^i_rQ$ to the $^0_0Q$. Breaking these symmetries
reduces the number of soft modes, see Sec.\ \ref{subsec:VI.B} below.

It is important to realize that the above $Q$--$Q$ correlation function
corresponds to a very general four--fermion correlation, with no
restrictions on the frequencies $n_1$ and $n_2$ other than that they must
have opposite signs. It is thus {\em not}
related to the fermion number density, or some other conserved quantity,
although the density--density correlation function can be obtained as
a linear combination of these more general propagators, see
Eq.\ (\ref{eq:3.32}). The physical reason for
the softness of these modes is thus not a conservation law, but rather a
spontaneously broken symmetry, viz. the rotation symmetry in $\psi$ or
$\eta$--space between fermion fields with positive and negative frequency
indices, respectively, or the symmetry between retarded and advanced
degrees of freedom. This symmetry is broken whenever one has a nonzero
density of states at the Fermi level (which is just the difference between
the spectra of the retarded and advanced Green functions, respectively),
and the soft $Q$ excitations are the corresponding Goldstone modes. They
are analogous to the transverse spin fluctuations in the classical
Heisenberg model.

\subsubsection{Separation of soft and massive modes, and the nonlinear 
                                                sigma--model for fermions}
\label{subsubsec:III.C.3}

We now know that the correlation functions of the $Q_{nm}$ with $nm<0$ are
soft, while those with $nm>0$ are massive. Our next goal is to separate
these degrees of freedom in such a way that the soft modes remain 
manifestly soft to all orders in perturbation theory, in analogy to the
treatment of the $O(N)$--symmetric Heisenberg model in 
Sec.\ \ref{subsubsec:III.C.1}. Such a separation was first achieved by
Sch{\"a}fer and Wegner\cite{SchaferWegner} for non--interacting electrons.
The generalization of their treatment for the case of interacting electron
was given in Ref.\ \onlinecite{fermions}. Here we sketch the logic of the 
reasoning, for details we refer to that reference.

The matrices $Q$ under consideration are complex $8Nn\times 8Nn$ matrices,
or, alternatively, quaternion--valued $4Nn\times 4Nn$ matrices. 
However, all of their
matrix elements are not independent; the definition of the $Q$ in terms
of the fermionic bispinors implies symmetry properties that reduce the
number of independent matrix elements. Using these symmetry properties,
one can show that the $Q$ can be diagonalized by means of elements of
the unitary symplectic group USp($8Nn,{\cal C}$).
\footnote[16]{This group is defined as the intersection of a unitary and a
symplectic group, ${\rm USp}(2n,{\cal C}) \equiv
{\rm U}(2n,{\cal C})\cap {\rm Sp}(2n,{\cal C})$\cite{Wybourne}. It can
be represented by unitary matrices that are also symplectic.} 
The most
general $Q$ can thus be written
\begin{equation}
Q = {\tilde {\cal S}}\ D\ {\tilde {\cal S}}^{-1}\quad,
\label{eq:3.34}
\end{equation}
where $D$ is diagonal, and ${\tilde {\cal S}}\in {\rm USp}(8Nn,{\cal C})$.

However, diagonalization is more than we want. Since we know that the
$Q_{nm}$ with $nm<0$ are soft, while those with $nm>0$ are massive,
we are interested in generating the most general $Q$ from a matrix $P$ that
is block--diagonal in Matsubara frequency space,
\begin{equation}
P = \left( \begin{array}{cc}
       P^> & 0   \\
       0   & P^< \\
    \end{array} \right)\quad,
\label{eq:3.35}
\end{equation}
where $P^>$ and $P^<$ are matrices with elements $P_{nm}$ where $n,m>0$ and
$n,m<0$, respectively. This can easily be achieved. Due to the symmetry
properties of $P^>$ and $P^<$,
the most general $P$ can be obtained from $D$ by an
element $U$ of ${\rm USp}(4Nn,{\cal C})\times {\rm USp}(4Nn,{\cal C})$.
The most general $Q$ can therefore be written
\begin{equation}
Q = {\cal S}\ P\ {\cal S}^{-1}\quad,
\label{eq:3.36}
\end{equation}
with ${\cal S}={\tilde {\cal S}}U^{-1}$. The set of transformations ${\cal S}$
is the set of
all cosets of ${\rm USp}(8Nn,{\cal C})$ with respect to
${\rm USp}(4Nn,{\cal C})\times {\rm USp}(4Nn,{\cal C})$, i.e. the
${\cal S}$ form the homogeneous space
${\rm USp}(8Nn,{\cal C})/{\rm USp}(4Nn,{\cal C})\times {\rm USp}(4Nn,{\cal C})$.

This achieves the desired separation of our degrees of freedom into soft
and massive ones. The massive degrees of freedom are represented by the
matrix $P$, while the soft ones are represented by the transformations
${\cal S}\in {\rm USp}(8Nn,{\cal C})/{\rm USp}(4Nn,{\cal C})\times
{\rm USp}(4Nn,{\cal C})$. The analogy with the $O(N)$--Heisenberg model is
now obvious: The unitary--symplectic coset space, identified above as the
space whose elements comprise the soft modes of the theory, is a
matrix generalization of the homogeneous space $O(N)/O(N-1)\times O(1)$,
which represents the soft modes ${\hat\phi}$ in the $O(N)$ vector model.

In order to formulate the field theory in terms of the soft and massive
modes, one also needs the invariant measure $I[P]$, or the Jacobian
of the transformation from the $Q$ to the $P$ and the $\cal S$, defined by
\begin{equation}
\int D[Q]\,\ldots = \int D[P]\,I[P] \int D[\cal S]\,\ldots\quad.
\label{eq:3.37}
\end{equation}
$I[P]$ has been constructed explicitly\cite{fermions}. However, it is not
necessary to know its explicit form to derive the nonlinear sigma--model.

The further development of the theory is analogous to the
derivation of Eqs.\ (\ref{eqs:3.24}), although the technical details
are more intricate\cite{fermions}. One splits the expectation value
$\langle P\rangle$ off the massive field $P$, and defines a new matrix
field
\footnote[17]{Strictly speaking, ${\hat Q}$ contains only the real part of
$\langle P\rangle$, and we also drop a normalization factor.}
\begin{equation}
{\hat Q}({\bf x}) = {\cal S}({\bf x})\,\langle P\rangle\,{\cal S}^{-1}({\bf x})
                                                                     \quad,
\label{eq:3.38}
\end{equation}
${\hat Q}$ has the properties ${\hat Q}^2 = {\rm const}$, and 
$\tr{\hat Q} = 0$.
The action can then be written as a
generalized matrix nonlinear sigma--model that
contains only the soft modes ${\hat Q}$, and corrections that are all
massive. The nonlinear sigma--model part of the action reads
\begin{mathletters}
\label{eqs:3.39}
\begin{equation}
{\cal A}_{NL\sigma M} = \frac{-1}{2G}\,\int d{\bf x}\
     \tr\left(\nabla\tilde Q ({\bf x})\right)^2
+ 2H \int d{\bf x}\ \tr\left(\Omega\,{\tilde Q}({\bf x})
     \right) + {\cal A}_{int}[\tilde Q]\quad,
\label{eq:3.39a}
\end{equation}
with the matrix field ${\tilde Q} = {\hat Q} - \langle P\rangle$, and
${\cal A}_{int}$ the interaction part of the action, Eq.\ (\ref{eq:3.5}),
replicated and rewritten in terms of the $Q$--matrices.
\footnote[18]{One can write either $\tilde Q$ and $\hat Q$ in the action,
Eq.\ (\ref{eq:3.39a}). The only resulting difference is an uninteresting
constant contribution to ${\cal A}_{NL\sigma M}$.}
The explicit derivation yields the
coupling constants as $G=8/\pi\sigma_0$, with $\sigma_0$ the conductivity
in self--consistent Born approximation, and $H=\pi N_F/8$ which can
be interpreted as the bare quasi--particle or specific heat density of
states\cite{CastellaniDiCastro,CastellaniKotliarLee,R}. $G$ serves as
a measure of the disorder in the system. $\Omega$ is a diagonal matrix
whose diagonal elements are the fermionic Matsubara frequencies $\omega_n$.
$\hat Q$ is subject to the constraints,
\begin{equation}
{\hat Q}^2({\bf x}) \equiv 1\quad,\quad
       {\hat Q}^{\dagger} = {\hat Q}\quad,\quad\tr{\hat Q}({\bf x}) = 0\quad.
\label{eq:3.39b}
\end{equation}
\end{mathletters}%

The effective action given by Eqs.\ (\ref{eqs:3.39}) is the generalized
nonlinear sigma--model that was first proposed by 
Finkel'stein\cite{Finkelstein,FinkelsteinZPhys} as a model 
for disordered interacting electrons,
and whose properties have been studied in considerable detail, in particular
with respect to the metal--insulator transition that is described by 
it\cite{Finkelstein2,CDLM,R,LandauMIT,opmit}. 
We will review the properties of this theory in
Sec.\ \ref{sec:VI} below.

\subsubsection{Order parameter field theory for metal--insulator transitions}
\label{subsubsec:III.C.4}

From Eq.\ (\ref{eq:3.33}) we know that the order parameter that is related
to the existence of Goldstone modes in the Fermi system is the density of
states at the Fermi level, $N(\epsilon_F)$. 
It is known that at the metal--insulator transition of interacting electrons,
$N(\epsilon_F)$ is critical\cite{R}. It is nonzero in the metallic phase, and
vanishes as the transition is approached. This suggests that $N(\epsilon_F)$
can be considered an order parameter for this type of transition, and that
one should try to derive an order parameter field theory for the
metal--insulator transition in analogy to those for the magnetic and
superconducting transitions discussed in Sec.\ \ref{subsec:III.B}. One way
of doing this is as follows\cite{LandauMIT,opmit}. 

In terms of the $Q$--matrix, $N(\epsilon_F)$ is given as the expectation
value of the diagonal matrix element $Q_{nn}$ in the limit of zero
frequency, see Eq.\ (\ref{eq:3.31}). A pure order parameter description would
therefore integrate out all modes except for $Q_{n=0,n=0}$.
Since the frequencies are coupled, however, it is adavantageous to keep all
of the modes that are massive in the ordered (i.e. metallic) phase, and to
integrate out the soft modes, i.e. the $Q_{nm}$ with $nm<0$. This can be
done in analogy to Eqs.\ (\ref{eqs:3.25}). We separate $\hat Q$ into
blocks in frequency space,
\begin{equation}
{\hat Q}_{nm} = \cases{Q_{nm}\quad, &if $n\geq 0, m\geq 0$ or $n<0$, $m<0$
                                                                     \quad,\cr
                       q_{nm}\quad, &if $n\geq 0$, $m<0$\quad,\cr
                       q^{\dagger}_{nm}\quad, &if $n<0$, $m\geq 0$\quad.\cr}
\label{eq:3.40}
\end{equation}
and enforce the constraint
${\hat Q}^2 = 1$, Eq.\ (\ref{eq:3.39b}), by means of a matrix Lagrange 
multiplier field $\Lambda$,
\begin{equation}
\prod_{\bf x}\ \delta\bigl[\hat Q^{2}({\bf x})-1\bigr] =
 \int D[\Lambda]\ \exp\Bigl\{-{1\over 2G}\int d{\bf x}\ \tr\ \bigl(\Lambda
 ({\bf x})[\hat Q^{2}({\bf x})-1]\bigr)\Bigr\}\quad,
\label{eq:3.41}
\end{equation}
where the factor of $1/2G$ has been inserted for convenience. By
decomposing $\Lambda$ into blocks like
$\hat Q$ one sees that the elements $\Lambda_{nm}$ with $nm>0$,
together with the tracelessness condition, Eq.\ (\ref{eq:3.39b}), are
sufficient for enforcing the constraint $\hat Q^{2}=1$. We can
therefore restrict ourselves to $\Lambda_{nm}$ with $nm>0$.
The action, Eq.\ (\ref{eq:3.39a}), can then be written
\begin{eqnarray}
{\cal A}_{NL\sigma M}[\hat{Q},\Lambda]=\frac{-1}{2G}\int d{\bf x}\ \tr
\biggl[\Lambda({\bf x})[\hat{Q}^2({\bf x})-
\openone]+\Bigl(\nabla\hat{Q}({\bf x})\Bigr)^2\biggr]+2H \int
d{\bf x}\ \tr\Bigl(\Omega \hat{Q}({\bf x})\Bigr)
+ {\cal A}_{int}[{\hat Q}]\quad.
\label{eq:3.42}
\end{eqnarray}
The soft modes $q$ can now be integrated out to obtain an effective
action entirely in terms of $Q$ and $\Lambda$. We obtain
\begin{eqnarray}
{\cal A}_{eff}&=&\ln\ \int D[q,q^{\dagger}]\ \exp\bigl({\cal A}_{NL\sigma M}
   [\tilde Q,\Lambda]\bigr)
\nonumber\\
&=&{\cal A}_{NL\sigma M}[Q,\Lambda] - {1\over 2}\ Tr\ \ln M[\Lambda]\quad.
\label{eq:3.43}
\end{eqnarray}
Here $M$ is a complicated matrix that depends linearly on $\Lambda$.
By constructing a saddle--point solution of ${\cal A}_{eff}$ one
can now construct a mean--field theory of the metal--insulator
transition in analogy to Eqs.\ (\ref{eq:3.26}). This will be discussed
in Sec.\ \ref{subsubsec:VI.B.2} below.

\section{Magnetic Transitions at Zero Temperature}
\label{sec:IV}

\subsection{Itinerant ferromagnets}
\label{subsec:IV.A}

Perhaps the most obvious example of a quantum phase transition is the
paramagnet--to--ferromagnet transition of itinerant electrons at $T=0$ as a
function of the exchange interaction between the electron spins. Early
theoretical work on this transition suggested that the critical behavior in
the physical dimensions $d=2$ and $d=3$ was mean--field like, as it is in the
thermal ferromagnetic transition in $d>4$. As discussed in Sec.\ \ref{sec:I}, 
the reason for this suggestion was that the 
phase transition in a quantum mechanical system
in $d$ dimensions is related to a classical phase transition in $d+z$
dimensions. Formally it was shown\cite{Hertz} that $z=3$ and $z=4$ in clean and
disordered itinerant quantum ferromagnets, respectively. This appeared to
reduce the upper critical dimension $d_c^{+}$, above which fluctuations are
unimportant and simple mean--field theory yields the correct critical
behavior, from $d_c^{+}=4$ in the classical case to $d_c^{+}=1$ and 
$d_c^{+}=0$ in the clean and disordered quantum cases, respectively. If this
were true, then this QPT would be uninteresting from a critical phenomena
point of view.
 
It is now known that this is not the case\cite{fmdirty,fmclean,fmsb}.
In the early work there was an implicit
assumption that the only modes or excitations that are important for
describing the phase transition were the critical modes. For the finite
temperature phase transitions this is generally correct.
\footnote[19]{However, complicated soft mode structures can occur
in classical systems as well, for instance in liquid crystals. See
Ref.\ \onlinecite{MazenkoRamaswamyToner} for an example.}
However, at zero temperature the statics and dynamics are coupled together,
which tends to increases the number of soft modes. Further, at $T=0$ there
are soft excitations that exist in the entire metallic phase due to a
spontaneously broken symmetry that is unrelated to magnetism. These
particle--hole excitations (`diffusons') and particle--particle 
excitations (`cooperons') were discussed in Sec.\ \ref{sec:III}. 
These modes cause what are known as weak localization effects,
\footnote[20]{The term `weak localization' is loosely defined, and often 
 misunderstood. We use it to refer to the
 nonanalytic behavior of electronic correlation functions in the limit
 of zero momentum and/or frequency that is induced by quenched disorder,
 or by a combination of interactions and quenched disorder, and that
 occurs even if the system is far from any kind of phase
 transition. The physical reason for these nonanalyticities is the diffusive
 motion of the electrons in the presence of quenched disorder. For a
 detailed discussion see, e.g., Refs.\ \onlinecite{AAKL},
 \onlinecite{LeeRamakrishnan}, and \onlinecite{ernst}.}
and they are soft due to their diffusive
nature. It turns out that these additional soft modes couple to the
critical modes and influence the critical behavior.

\subsubsection{Disordered ferromagnets}
\label{subsubsec:IV.A.1}

We now proceed with the discussion of the LGW functional derived in
Sec.\ \ref{subsec:III.B}, along the lines of Ref.\ \onlinecite{fmdirty}.
As discussed in Sec.\ \ref{subsec:III.B},
formally expanding the LGW functional $\Phi[{\bf M}]$, Eq.\ (\ref{eq:3.12b}),
in powers of ${\bf M}$ yields Eq.\ (\ref{eq:3.12c}). Of particular
interest to us is the quadratic or Gaussian term, which in Fourier space 
is given by
\begin{equation}
\Phi_2({\bf M}) = \frac{1}{2}\sum_{{\bf q},\Omega_n}\sum_{\alpha}\ 
   \left[1/\Gamma^{(t)} - \chi_s({\bf q},\Omega_n)\right]\,
          \vert {\bf M}^{\alpha}({\bf q},\Omega_n)\vert^2\quad.  
\label{eq:4.1}
\end{equation}
Here we have scaled ${\bf M}$ with $1/\Gamma^{(t)}$. $\chi_s$ is the Fourier transform of the dynamical spin
susceptibility in the reference ensemble,
\begin{mathletters}
\label{eqs:4.2}
\begin{equation}
\chi_s(x_1-x_2)=\left\langle n_{s,a}^{\alpha}(x_1)n_{s,a}^{\alpha}
(x_2)\right\rangle_{ref}\quad,  
\label{eq:4.2a}
\end{equation}
where $n_{s,a}\ (a=1,2,3)$ is one component of the spin density
vector ${\bf n}_s$, Eq.\ (\ref{eq:3.4b}).
Spin density conservation implies that at small frequency and
wavenumber, the Fourier transform of $\chi_s$ has a diffusive 
structure\cite{Forster},
\begin{equation}
\chi_s({\bf q},\Omega_n) = \chi_0({\bf q})\,\frac{D{\bf q}^2}{\vert\Omega_n
                           \vert + D{\bf q}^2}\quad,
\label{eq:4.2b}
\end{equation}
where $D$ is the spin diffusion coefficient and $\chi_0({\bf q})$ is the
static spin susceptibility, both in the reference ensemble. In order to reach
criticality, the frequency must be taken to zero before the wavenumber.
\footnote[21]{This can be seen as follows. Since the magnetization
 is conserved, ordering on a length scale $L$ requires some spin density to be
 transported over that length, which takes a time $t\sim L^2/D$, with $D$
 the spin diffusion coefficient. Now suppose the coherence length is $\xi$, 
 and we look
 at the system at a momentum scale $q$ or a length scale
 $L\sim 1/q < \xi$. Because of the time it takes the system to order over
 that scale, the condition for criticality is $L^2 < {\rm Min} (D t,\xi^2)$.
 In particular, one must have $L^2 < Dt$, or $\Omega \sim 1/t < Dq^2$.}
In the critical limit, we can thus expand,
\begin{equation}
\chi_s({\bf q},\Omega_n) = \chi_0({\bf q})\,\bigl[1 - \vert\Omega_n\vert
           /D{\bf q}^2 + \ldots\bigr]\quad. 
\label{eq:4.2c}
\end{equation}
\end{mathletters}%
It can be easily verified that the corrections to the leading
terms in Eq.\ (\ref{eq:4.2c}) are irrelevant for the critical behavior.

Next comes the crucial step of calculating $\chi_0({\bf q})$ in the 
reference ensemble. For a long time it was assumed that $\chi_0({\bf q})$ 
was an analytic function of ${\bf q}^2$. This might seem plausible, given
that the reference ensemble describes a physical system that is far away 
from any critical point. However, it is now known that this reasoning
is incorrect, and that any itinerant electron system at zero temperature
has long--range correlations everywhere in its phase diagram, even
far away from any critical point\cite{ernst}.
These long--range correlations are due to the extra soft
modes that generically exist in quantum systems. In disordered systems,
these effects are known collectively as weak localization effects.
For small values of $\vert{\bf q}\vert$, and for $d>2$, the leading behavior
of $\chi_0({\bf q})$ is
\begin{equation}
\chi_0({\bf q}) = c_0 - c_{d-2}\vert{\bf q}\vert^{d-2} - c_2{\bf q}^2\quad,
\label{eq:4.3}
\end{equation}
where the $c_i$ are positive constants.
Notice that the susceptibility decreases with increasing wavenumber.
The negative sign of the $\vert{\bf q}\vert^{d-2}$ term, whose prefactor
$c_{d-2}$ vanishes for noninteracting electrons, is due to the fact that
the interaction increases $\chi_0$, and its effect gets weaker at larger
values of $\vert{\bf q}\vert$.
For $d<2$, the electrons are localized and a different theory is needed.

Using Eqs.\ (\ref{eqs:4.2}) and (\ref{eq:4.3}), the Gaussian part of the 
LGW functional can be written more explicitly,
\begin{mathletters}
\label{eqs:4.4}
\begin{equation}
\Phi_2({\bf M}) = \frac{1}{2}\sum_{{\bf q},\Omega_n}\sum_{\alpha}\,
                   \bigl[ t_0 + {\bf q}^{d-2} + {\bf q}^2 + \vert\Omega_n\vert
                    /{\bf q}^2\bigr]\vert {\bf M}^{\alpha}({\bf q},\Omega_n)
                         \vert^2 \quad,  
\label{eq:4.4a}
\end{equation}
where we have omitted the prefactors of the various terms, and
\begin{equation}
t_0 = \frac{1}{\Gamma^{(t)}} - \chi_0({\bf q}=0)\quad,  
\label{eq:4.4b}
\end{equation}
\end{mathletters}%
is the bare distance from the critical point. We recognize 
Eq.\ (\ref{eq:4.4b}) as a generalized Stoner criterion for ferromagnetism. 
Physically, the most important term in Eq.\ (\ref{eq:4.4a}) is the 
$\vert{\bf q}\vert^{d-2}$, which means that in real space
the spin density fluctuations interact through a potential that 
decays with distance $r$ as $r^{2-2d}$. 
It is well known from the theory of classical 
phase transitions
that long--range interactions suppress fluctuation effects, and the
critical behavior in systems with such interactions can therefore be 
determined exactly\cite{FisherMaNickel}. 
For example, using renormalization group
methods it can be established that all terms higher than $l=2$ in 
the Landau expansion, Eq.\ (\ref{eq:3.12c}),
are renormalization group irrelevant for $d>2$, so that the 
upper critical dimension for
fluctuations is $d_c^+=2$. From this it follows that many of the
critical exponents can be determined exactly from $\Phi_2$.

To see this we define the order parameter correlation function by
\begin{mathletters}
\label{eqs:4.5}
\begin{equation}
G({\bf q},\Omega_n) = \left\langle\vert M_a({\bf q},\Omega_n)\vert^2
                        \right\rangle \quad,
\label{eq:4.5a}
\end{equation}
with $M_a$ an arbitrary component of the vector field ${\bf M}$.
For small $\vert{\bf q}\vert$ and $\Omega_n$ we find
\begin{equation}
G({\bf q},\Omega_n) = \frac{1}{t + \vert{\bf q}\vert^{d-2} + {\bf q}^2 
                                         + \Omega_n/{\bf q}^2}\quad.
\label{eq:4.5b}
\end{equation}
\end{mathletters}%
Here we have again omitted all prefactors of the various terms in 
the denominator in
Eq.\ (\ref{eq:4.5b}), since they are irrelevant for our purposes. 
In this equation we have
also anticipated that irrelevant variables will renormalize the $t_0$ in 
Eq.\ (\ref{eq:4.4a}) to $t$, the actual distance from the critical point, 
in Eq.\ ({\ref{eq:4.5b}). For 
${\bf q} = \Omega_n = 0$, the correlation function $G$ determines the
magnetic susceptibility $\chi_m\sim G({\bf q}=0,\Omega=0)$ in
zero field. Hence we have $\chi_m(t)\sim t^{-1}=t^{-\gamma}$, where the
last relation defines the critical exponent $\gamma$. In our case,
\begin{equation}
\gamma =1\quad,  
\label{eq:4.6}
\end{equation}
which is valid for all $d>2$. $\gamma$ thus has its usual
mean--field value. However, for non--zero wavenumbers 
the anomalous $\vert{\bf q}\vert^{d-2}$ term
dominates the usual ${\bf q}^2$ dependence for all $d<4$. By inspection, the
correlation length exponent, defined by $\xi \sim t^{-\nu}$, is given by
\begin{equation}
\nu = \cases{1/(d-2)\quad,&for $2<d<4$\quad,\cr 
               1/2\quad,  &for $d>4$\quad.\cr}
\label{eq:4.7}
\end{equation}
Note that $\nu \geq 2/d$, as it must be according to the
discussion in Sec.\ \ref{sec:I}. The wavenumber dependence of $G$ at 
$t=0$ is
characterized by the exponent $\eta$, which is defined as 
$G({\bf q},\Omega_n = 0)\sim \vert{\bf q}\vert^{-2+\eta}$. 
From Eq.\ (\ref{eq:4.5b}) we obtain
\begin{equation}
\eta = \cases{4-d\quad,&for $2<d<4$\quad,\cr
               0 \quad,&for $d>4$\quad.\cr}
\label{eq:4.8}
\end{equation}
Finally, the dynamical scaling exponent $z$ is read off the relation
$\xi ^z\sim \xi ^2/t$ to be
\begin{equation}
z = \cases{d\quad, &for $2<d<4$\quad,\cr 
           4\quad, &for $d>4$\quad.\cr}
\label{eq:4.9}
\end{equation}
Note that all of these exponents `lock into' their mean--field values
at a dimensionality $d_c^{++}=4$.

To determine the critical exponents $\beta$ and $\delta$ we need the
equation of state in the ordered phase. An important point is that although
the quartic ($l=4$) term is irrelevant in the technical sense of the
renormalization group,
it is nevertheless needed to determine the critical behavior of the 
magnetization. The reason is that the coefficient of
the ${\bf M}^4$--term is an example of a dangerous irrelevant
variable. Further, the same mechanism that causes the nonanalyticity in 
Eq.\ (\ref{eq:4.3})
leads to a divergence of that coefficient in the long--wavelength,
low--frequency limit, i.e. the function $X^{(4)}$ in Eq.\ (\ref{eq:3.12c})
cannot be localized in space and time. The terms with
$l>4$ in Eq.\ (\ref{eq:3.12c}) are even more singular in this limit.
Despite this unpleasant behavior of the field theory, it is easy to see by
power counting that all of these terms are irrelevant with
respect to the Gaussian fixed
point. Taking these singularities into account in constructing the equation
of state gives for $d>2$,
\begin{equation}
tm + m^{d/2} + m^3 = h\quad,  
\label{eq:4.10}
\end{equation}
where we again have left off the irrelevant prefactors. Here $m$ is the
modulus of the magnetization vector, and $h$ is the external magnetic
field. Notice the
term $m^{d/2}$, which occurs in addition to what is otherwise an ordinary
mean--field equation of state. It occurs because of the singularities
mentioned above. For $d<6$ it dominates the $m^3$ term, and hence determines the
critical exponents $\beta$ and $\delta$. The zero--field magnetization near
criticality behaves by definition of the exponent $\beta$ as 
$m(t,h=0)\sim t^\beta$. The equation of state, Eq.\ (\ref{eq:4.10}), gives
\begin{mathletters}
\label{eqs:4.11}
\begin{equation}
\beta = \cases{2/(d-2)\quad &for $2<d<6$\quad,\cr 
               1/2 \quad &for $d>6$\quad.\cr}
\label{4.11a}
\end{equation}
Similarly, the exponent $\delta$, defined by $m(t=0,h)\sim h^{1/\delta}$, is
\begin{equation}
\delta  = \cases{d/2\quad,&for $2<d<6$\quad,\cr 
                   3\quad,&for $d>6$\quad.\cr}
\label{eq:4.11b}
\end{equation}
\end{mathletters}%
Note that these relations imply yet another upper critical
dimension, $d_c^{+++}=6$, defined as the dimension where $\beta$ and
$\delta$ `lock into' their mean--field values of $1/2$ and $3$, 
respectively.

The critical behavior of the specific heat, $c_V$, has also been
calculated. It is most convenient to discuss the specific heat coefficient, 
$\gamma_V = \lim_{T\rightarrow 0}c_V/T$, which in a Fermi liquid would
be a constant. Its behavior at criticality, $t=0$, is adequately
represented by the following integral,
\begin{equation}
\gamma_V = \int_0^{\Lambda}dq\ \frac{q^{d-1}}{T+q^d+q^4+h^{1-1/\delta }/q^2}
                                                                       \quad. 
\label{eq:4.12}
\end{equation}
Remarkably, in zero magnetic field, $\gamma_V$ diverges
logarithmically as $T\rightarrow 0$ for all dimensions $2<d<4$. From a
scaling viewpoint this is a consequence of the dynamical scaling exponent $z$
being exactly equal to the spatial dimensionality, $d$, in this dimensionality
range. In terms of a scaling function, Eq.\ (\ref{eq:4.12}) implies
\begin{equation}
\gamma_V(t,T,h)=\Theta (4-d)\ln b + F_{\gamma}(tb^{1/\nu},Tb^z,hb^{\delta
\beta /\nu })\quad.  
\label{eq:4.13}
\end{equation}
Here $\Theta (x)$ denotes the step function, $b$ is an arbitrary
scale factor, and $F_{\gamma}$ is a scaling function that cannot be determined
by general arguments alone.

The complexity of the critical behavior displayed above is astonishing,
given that it is controlled, after all, by a simple Gaussian fixed point. 
For instance, the appearance of several different `upper critical 
dimensionalities' is suprising at first sight. As the scaling analysis
makes clear, this is a direct consequence of the nonanalytic terms in the
LGW functional, which in turn results from the `additional' soft modes.
Furthermore, the critical behavior of different quantities can be
affected by different dangerous irrelevant variables. For instance,
the critical behavior of the magnetization is governed by the coefficient
of the (irrelevant) quartic term in the LGW functional, while the specific
heat coefficient is not affected by this dangerous irrelevant variable.
As a result, the Wilson ratio, $W = (m/h)/\gamma_V$, diverges at the
criticality, while general scaling\cite{SachdevZPhys} predicts that
it should be a universal finite number.

Another interesting example for the breakdown of general scaling is
provided by the temperature dependence of the magnetization $m$, and of
the magnetic susceptibility $\chi_m$. A detailed analysis\cite{fmdirty,fmclean}
reveals that these quantities do not depend on the critical temperature
scale, and therefore their temperature scaling behavior is not determined
by the dynamical critical exponent $z$, but rather by the scale dimensions
of subleading temperature scales. One finds for $m$ and $\chi_m$ the
following homogeneity laws,
\begin{mathletters}
\label{eqs:4.14}
\begin{equation}
m(t,T,H) = b^{-\beta/\nu} m(tb^{1/\nu}, Tb^{\phi/\nu}, Hb^{\delta\beta/\nu})
                                                                    \quad,
\label{eq:4.14a}
\end{equation}
\begin{equation}
\chi_m(t,T,H) = b^{\gamma/\nu} \chi_m(tb^{1/\nu}, Tb^{\phi/\nu},
                           Hb^{\delta\beta/\nu})\quad.
\label{eq:4.14b}
\end{equation}
Here the crossover exponent $\phi$ turns out to be given {\em not} by $\nu z$,
but rather by\cite{fmclean},
\footnote[22]{Refs.\ \onlinecite{fmdirty,fmsb} contained a mistake in the
result for $\phi$, which was corrected in Ref.\ \onlinecite{fmclean}.}
\begin{equation}
\phi = \cases{2\nu = 2/(d-2)\quad, &for\quad $2<d<\sqrt{5}+1$\quad,\cr
              d/2\quad,            &for\quad $\sqrt{5}+1<d<4$\quad,\cr
              4/(d-2)\quad,        &for\quad $4<d<6$\quad,\cr
              2\nu = 1\quad,       &for\quad $d>6$\quad.\cr}
\label{eq:4.14c}
\end{equation}
\end{mathletters}%
The correct physical interpretation of the temperature dependence of $m$ and
$\chi_m$ is not dynamical scaling, but rather
the crossover from the quantum critical region to a regime whose
behavior is dominated by the classical Gaussian fixed
point\cite{Millis}.

\subsubsection{Clean ferromagnets}
\label{subsubsec:IV.A.2}

In the previous subsection, the problem of disordered ferromagnets was
considered. The only point in that dicussion where the disorder was
important was the diffusive dispersion relation of the `extra' soft
modes. This raises the question whether similar effects might
exist in clean itinerant ferromagnets. The first question that arises
in the context is what, if anything, will replace the ${\bf q}^{d-2}$--term
in the static spin susceptibility, Eq.\ (\ref{eq:4.3}), in clean systems.
To answer this, let us consider the perturbation theory for 
$\chi_0({\bf q})$.
What leads to the nonanalyticity in Eq.\ (\ref{eq:4.3}) is the coupling of
two diffusive modes, which mathematically takes the form of a mode--mode
coupling integral of the type
\begin{equation}
\int d{\bf k}\int d\omega\ \frac{1}{\omega + {\bf k}^2}\,\frac{1}
          {\omega +({\bf q} + {\bf k})^2}\quad,  
\label{eq:4.15}
\end{equation}
with ${\bf q}$ the external momentum, and we have set the diffusion
coefficient equal to unity. Renormalization group techniques have
been used to show
that this is indeed the leading small--${\bf q}$ behavior of $\chi_0$.

What changes in a clean system? The soft modes are still the density and spin
density fluctuations, and in addition more general particle--hole excitations. 
All of these have a linear dispersion relation, i.e., 
$\omega \sim \vert{\bf q}\vert$. One might thus expect $\chi_0({\bf q})$ 
in a clean
system to have a mode--mode coupling contribution analogous to that given by
Eq. (\ref{eq:4.15}), but with ballistic modes instead of diffusive ones,
\begin{equation}
\int d{\bf k}\int d\omega\ \frac{1}{\omega +\vert{\bf k}\vert}\,\frac
                        {1}{\omega +\vert{\bf q}+{\bf k}\vert}\quad.  
\label{eq:4.16}
\end{equation}
In generic dimensions, expanding in $\vert{\bf q}\vert$ leads 
to\cite{fmclean},
\begin{equation}
\chi_0({\bf q})\sim {\rm const} + d_{d-1} \vert{\bf q}\vert^{d-1}
                                + d_{2}\vert{\bf q}\vert^2\quad,  
\label{eq:4.17}
\end{equation}
at $T=0$. Here $d_{d-1}$ and $d_2$ are constant prefactors.
For $d\leq 3$, the nonanalytic term in Eq.\ (\ref{eq:4.17})
represents the leading ${\bf q}$--dependence of $\chi_0$. In $d=3$,
one finds a ${\bf q}^2\ln 1/\vert{\bf q}\vert$ term, and in $d>3$ the
analytic contribution is the leading one.

If $d_{d-1}<0$ in Eq.\ (\ref{eq:4.17}), 
then the same arguments used in the
previous section apply here. For $d<3$ one obtains the critical exponents
\begin{equation}
\beta =\nu =1/(d-1)\quad,\quad\eta =3-d\quad,\quad\delta=z=d\quad,\quad\gamma=1
                                                               \quad,  
\label{eq:4.18}
\end{equation}
and Eq.\ (\ref{eq:4.13}) is valid with $4-d$ replaced by $3-d$. For $d>3$
all of the exponents have their mean--field values, and in $d=3$ there are
logarithmic corrections to mean--field like critical behavior.

As in the disordered case discussed in the previous subsection, the
temperature scaling of the magnetization is complicated and violates
general scaling. The homogeneity laws for $m$ and $\chi_m$ are given
by Eqs.\ (\ref{eq:4.14a}), (\ref{eq:4.14b}), 
with a crossover exponent $\phi$ given by\onlinecite{fmclean}
\begin{equation}
\phi = \cases{\nu = 1/(d-1)\quad, &for $1<d<2$\quad,\cr
              d/2(d-1)\quad,      &for $2<d<3$\quad,\cr
              3/(d+1)\quad,       &for $3<d<5$\quad,\cr
              \nu = 1/2\quad,     &for $d>5$\quad.\cr}
\label{eq:4.19}
\end{equation}

These results for the quantum critical behavior of clean itinerant ferromagnets
disagree with those of Ref.\ \onlinecite{Hertz}. It should be emphasized that
these discrepancies cannot be related to differences in the underlying models.
If the model of Ref.\ \onlinecite{Hertz} is properly renormalized, then it
shows the same critical behavior as discussed above, since all features of
the action that are not included in the bare model will be generated by the
renormalization group. This was shown in
Ref.\ \onlinecite{fmclean}. It highlights the important point that a
simple power counting analysis yields the right answer only if no qualitatively
new terms are generated under renormalization.

Detailed calculations have confirmed the existence of the 
$\vert{\bf q}\vert^{d-1}$--term in Eq.\ (\ref{eq:4.17})\cite{chi_s}.
However, to lowest order in perturbation theory the coefficient $d_{d-1}$
of that term is positive. For $d\leq 3$ this implies that $\chi_0$
increases with increasing ${\bf q}$ like $\vert{\bf q}\vert^{d-1}$. 
For any physical system for which this were the true asymptotic behavior 
at small ${\bf q}$, this would have
remarkable consequences for the zero--temperature phase transition from the
paramagnetic to the ferromagnetic state as a function of the exchange
coupling. One possibility is that the ground state of the system will not be
ferromagnetic, irrespective of the strength of the spin triplet interaction,
since the functional form of $\chi_0$ leads to the instability of any
homogeneously magnetized ground state. Instead, with increasing interaction
strength, the system would undergo a transition from a paramagnetic Fermi
liquid to some other type of magnetically ordered state, most likely a spin
density wave. While there seems to be no observational evidence for this,
let us point out that in $d=3$ the effect is only logarithmic, and would
hence manifest itself only as a phase transition at exponentially small
temperatures, and exponentially large length scales, that might well be
unobservable. For $d\leq 2$, on the other hand, there is no long--range
Heisenberg ferromagnetic order at finite temperatures, and the suggestion
seems less exotic. Furthermore, any finite concentration of quenched
impurities will reverse the sign of the leading nonanalyticity, and thus
make a ferromagnetic ground state possible again.

Another possibility is that the zero--temperature paramagnet--to--ferromagnet
transition is of first order. It has been shown in Ref. \onlinecite{fmclean}
that the
nonanalyticity in $\chi_0({\bf q})$ leads to a similar nonanalyticity in the
magnetic equation of state, which takes the form
\begin{equation}
tm - v_d\,m^d + u\,m^3 = h\quad.  
\label{eq:4.20}
\end{equation}
with $m$ the magnetization, $h$ the external magnetic field, and $u>0$ 
a positive coefficient. If the soft mode mechanism discussed above is
the only mechanism that leads to nonanalyticities, then the sign of the
remaining coefficient $v_d$ in Eq.\ (\ref{eq:4.20}) should be the same 
as that of $d_{d-1}$ in Eq.\ (\ref{eq:4.17}), i.e. $v_d>0$. 
This would imply a first order
transition for $2<d<3$. In this case the length scale that the scenario
of the previous
paragraph would have been attributed to a spin density wave would instead be
related to the critical radius for nucleation at the first order phase
transition.

Finally, it is possible that terms of higher order in the interaction
amplitude could
change the sign of $d_{d-1}$ in Eq.\ (\ref{eq:4.17}) 
for interaction strengths
sufficient to cause ferromagnetism. In this case one obtains a continuous
quantum phase transition with critical
exponents as given by Eqs.\ (\ref{eq:4.18}) and (\ref{eq:4.19}).
Further work is clearly needed on the clean itinerant quantum ferromagnetism
problem.

\subsection{Disordered antiferromagnets}
\label{subsec:IV.B}

Quantum antiferromagnetism (AFM) has experienced a surge of interest in
recent years, mainly in efforts to explain the magnetic properties of
high--$T_c$ materials. Much of the theoretical
work has focused on one--dimensional spin chains, both with and without
quenched disorder. In $d=3$, the QPT in clean itinerant systems is 
mean--field like, since $z=2$ and hence $d+z$
is greater than the classical upper critical dimension $d_c^+=4$ for all
$d>2$. In this case Hertz's LGW theory\cite{Hertz} does {\em not} break
down, since the order parameter (the staggered magnetization) is not a
homogeneous quantity. Therefore the additional soft modes that make
the ferromagnetic problem so interesting do not couple to the order
parameter fluctuations, and the vertices
in Eq.\ (\ref{eq:3.12c}) can be localized in space and time. 
There has been relatively little work, on the other hand, on the 
disordered AFM problem in $d>1$. Even though in this case as well the soft
noncritical modes of the previous section do not influence the AFM
transition, the inclusion of quenched disorder leads to a very interesting
and difficult problem.

Following earlier work by Das Gupta and Ma\cite{DasGuptaMa}, 
Bhatt and Lee\cite{BhattLee} have studied
some aspects of the interplay between AFM and strong quenched disorder. They
considered a model for the insulating phase of a doped semiconductor
consisting of an ensemble of randomly distributed, AFM coupled Heisenberg
spins with a very broad distribution of coupling constants, $J$. Using a
numerical renormalization procedure they found that with 
decreasing temperature, an
increasing number of spin pairs freeze into inert singlets, but that some
unpaired spins remain and give an essentially 
free spin contribution, $\chi_m\sim 1/T$, to the magnetic
susceptibility as $T\rightarrow 0$. The net result was a
`random--singlet' phase, with a sub--Curie power law $T$ dependence of
the magnetic susceptibility as $T\rightarrow 0$. The quantum nature of the
spins thus prevents the classically expected long--range order of
either AFM or spin glass type.

Bhatt and Fisher\cite{BhattFisher} 
applied similar ideas to the disordered metallic phase.
They argued that rare fluctuations in the random potential always lead to
randomly distributed local moments to which the methods of Bhatt and
Lee can be applied, but now in the metallic phase. Their conclusion was that
the local moments cannot be quenched by either the Kondo effect or the 
RKKY interaction induced by the conduction electrons. 
This again leads to a random--singlet phase with a
magnetic susceptibility that diverges as $T\rightarrow 0$, albeit slower
than any power.

These results seem to imply the unlikely conclusion that antiferromagnetic 
long--range order can never exist in the ground state of disordered systems. 
In $d=1$ this conclusion is known to be wrong\cite{DSFrandomafm}. 
In higher dimensions it clearly
warranted further investigation. In Ref.\ \onlinecite{afm} it was established
that in some parts of the phase diagram, long--range order can exist. 
The basic idea is that quantum
fluctuations weaken the metallic random--singlet phase since they enhance the
interaction of the isolated local moment electrons with their environment. 
These interactions in turn restore long--range order by suppressing the 
random--singlet phase that would otherwise pre--empt an AFM transition.

This result was established by studying a model of an itinerant AFM with a
spatially random N{\'e}el temperature, or mass term. This model can be derived
from a fermionic description,
\footnote[23]{Of course it is necessary to consider fermions on a lattice, as a
continuum model does not display antiferromagnetism.}
just like the ferromagnetic one was derived in Sec.\ \ref{sec:III}.
The only difference is that the antiferromagnetic
order parameter, ${\bf N}({\bf x},\tau)$, is multiplied by a
phase function that represents perfect AFM order. Because of this phase
factor the singular soft modes responsible for the nonlocalities in the
ferromagnetic LGW theory do not couple to the order parameter. 
The resulting replicated local 
field theory for ${\bf N}$ has an action
\begin{eqnarray}
S[{\bf N}^{\alpha}]&=&{1\over 2}
                       \int d{\bf x}\,d{\bf y}\,\int_0^{1/T} d\tau\,d\tau'\,
                       \sum_{\alpha}\
                       {\bf N}^{\alpha}({\bf x},\tau)\,
                       \Gamma({\bf x}-{\bf y},
                       \tau - \tau')\,{\bf N}^{\alpha}({\bf y},\tau')
\nonumber\\
 &&+ u\int d{\bf x}\,\int_0^{1/T} d\tau\,\sum_{\alpha}\left(
  {\bf N}^{\alpha}({\bf x},\tau)\cdot {\bf N}^{\alpha}({\bf x},\tau)\right)^2
\nonumber\\
 &&-\Delta\int d{\bf x}\,\int_0^{1/T} d\tau\,d\tau'\,\sum_{\alpha,\beta}\ \left(
  {\bf N}^{\alpha}({\bf x},\tau)\cdot {\bf N}^{\alpha}({\bf x},\tau)\right)\,
  \left({\bf N}^{\beta}({\bf x},\tau')\cdot
         {\bf N}^{\beta}({\bf x},\tau')\right)\quad.
\label{eq:4.21}
\end{eqnarray}
The Fourier transform of the bare two--point vertex function 
$\Gamma ({\bf x},\tau)$ is
\begin{equation}
\Gamma ({\bf q},\Omega_n) = t + {\bf q}^2 + \vert\Omega_n\vert\quad.
\label{eq:4.22}
\end{equation}
$u$ is the coupling constant for the usual quartic term, and
$\Delta$ is the strength of the random mass term. Physically, $u$ is a 
measure of the strength
of quantum fluctuation effects, while $\Delta$ is a measure of disorder
fluctuations. Note that $u,\Delta\geq 0$, so the presence of disorder
(i.e., $\Delta\neq 0)$ has a destabilizing effect on the field theory.
To initiate the analysis of Eq.\ (\ref{eq:4.21}) we put $T=0$, and first
consider the case $\Delta =0$. We again define the scale dimension of a length
$L$ to be $[L]=-1$ (see Eq.\ (\ref{eq:2.5})),
and that of time to be $[\tau]=-z$, and look for a Gaussian fixed
point where $z=2$ and $\eta =0$. Power counting in $d$ dimensions shows
that $[u]=4-(d+z)$, so that $u$ is irrelevant, and the Gaussian fixed point
is stable,
for all $d>2$. The other exponents can then be readily determined; 
for instance, $\nu =1/2$. In contrast, the term proportional to $\Delta$
carries an extra time integral, so with respect to the Gaussian fixed
point we have 
$[\Delta] = 4-d$. Hence the disorder is relevant for $d<4$, and the Gaussian
fixed point
is no longer stable in the presence of disorder. This instability of the
Gaussian fixed point 
can also be inferred from the Harris criterion\cite{Harris}.

In order to see if there are other fixed points that might be stable,
a one--loop renormalization group calculation for the model given by
Eq.\ (\ref{eq:4.21}) was performed in Ref.\ \onlinecite{afm}.
The results are shown in a schematic flow diagram on the critical surface
in Fig.\ \ref{fig:4.1}.
\begin{figure}[t]
\centerline{\psfig{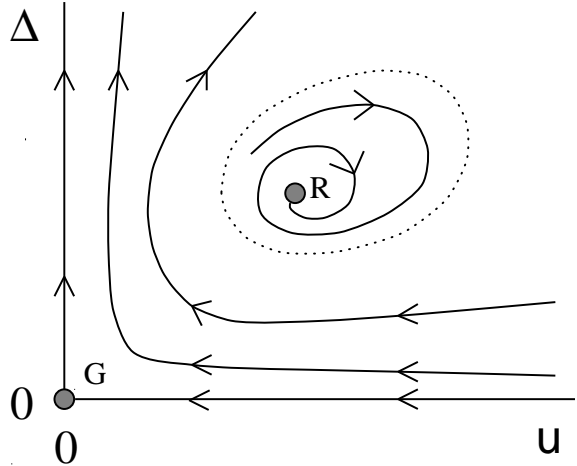}\vspace*{5mm}}
\caption{Schematic flow diagram on the critical surface. The unstable
 Gaussian (G) and the stable random (R) fixed points are shown. The
 dotted line denotes the boundary of the basin of attraction for the
 random fixed point. (From Ref.\ \protect\onlinecite{afm}.)}
\label{fig:4.1}
\bigskip
\end{figure}
One finds indeed a stable, critical fixed point that we call the random
fixed point. In Figure\ \ref{fig:4.1}, both the Gaussian and the random
fixed points are indicated. The linear renormalization group
eigenvalues on the critical surface are complex, so that the corrections to
scaling at the transition corresponding to the random 
fixed point are 
oscillatory in nature. More importantly, the random fixed point 
has only a limited range of
attraction. If the initial values of $\Delta$ and $u$ are
inside the basin of attraction for that fixed point,  
then the system undergoes a continuous phase
transition, and standard techniques can be used to calculate the critical
exponents. If,
on the other hand, the initial values of $\Delta $ and $u$ place the 
system outside of that basin of attraction, then
one finds runaway flows. The correct interpretation
of this result is not obvious. One possible interpretation is a fluctuation
induced first order phase transition. However, for a state with simple
AFM order this possibility is not realized\cite{afmRajesh}.
\begin{figure}[hb]
\centerline{\psfig{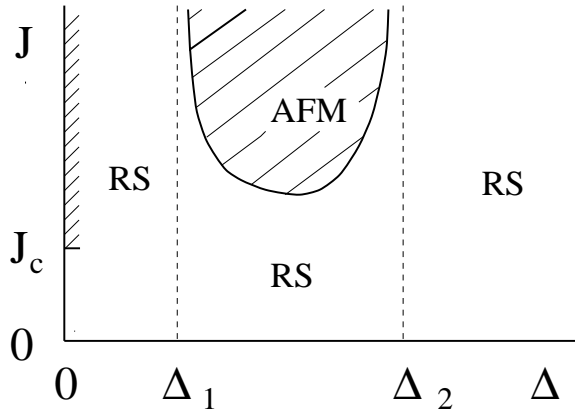}\vspace*{5mm}}
\caption{Schematic phase diagram,
 showing the antiferromagnetic
 (AFM) and random--singlet (RS) phases in the $J - \Delta$
 plane, with $J$ the effective AFM coupling constant, and $\Delta$ from
 Eq.\ (\protect\ref{eq:4.21}). For $\Delta =0$, one has an AFM for $J>J_c$, 
 and a Fermi
 liquid for $J<J_c$. The dependence of the phase boundary on the fluctuation
 parameter $u$ is not shown. (From Ref.\ \protect\onlinecite{afm}.)}
\label{fig:4.2}
\bigskip
\end{figure}
Another obvious possibility is that for
some parameter values, long--range order is not possible.
The latter possibility is consistent with the phase diagram shown in 
Fig.\ \ref{fig:4.2}, in which the random--singlet phase (no long--range order) 
and the AFM phase compete with one another. 
The basic idea is that although disorder destabilizes the field theory, and
hence works against ordering, quantum fluctuations as measured by $u$
can counteract this and lead to long--range order. For this to happen, however,
the initial values at $u$ and $\Delta $ must not be too small since $u$ is a
renormalization group irrelevant variable for $d=4-\epsilon$. 
In other regions of parameter space where no critical fixed point exists, 
the disorder wins out and scales to large
values. The resulting field theory is unstable, but has local instanton or
local magnetic moment solutions. Presumably these local solutions represent
the random singlet phase. Qualitatively, the most striking prediction 
is that for small disorder, there is no long--range order, 
but that for intermediate values of $\Delta$,
long--range order does exist. For further discussions we refer 
to the original publication\cite{afm}.

\section{Superconductor--Metal Transition at Zero Temperature}
\label{sec:V}
 
The transition from a normal metal to a superconductor at $T=0$ provides
an interesting contrast to the magnetic transitions discussed in
Sec.\ \ref{sec:IV}\cite{sc_letter}.
Again there are soft modes in addition to the order
parameter fluctuations, in this case particle--hole excitations in the
Cooper channel or `cooperons', that are integrated out in deriving a 
LGW functional, and that lead to long--range effective interactions between 
the order parameter fluctuations. For superconductors, 
the additional soft modes
are stronger than in the magnetic case, and they completely dominate the
physics. As a result, the quantum critical behavior of superconductors is
BCS--like. There are, however, very strong corrections to scaling that lead
to a broad transition region as $d\rightarrow 2$.

Let us concentrate on the realistic case of a superconductor with a
nonzero density of elastic scatterers, so that the
normal state conductivity is finite. The transition can then be triggered
by varying either the attractive electron--electron interaction, or the 
disorder. We proceed by performing 
a Landau expansion or expansion of the LGW functional $\Phi$,
Eq.\ (\ref{eq:3.14}), in powers of the order parameter. 
Since gauge invariance is not broken in the reference ensemble,
only even powers of $\Psi$ appear. The coefficients in this expansion,
i.e. the vertex functions of the effective field theory, are connected
correlation functions of the anomalous or Cooper channel density $n_c$,
Eq.\ (\ref{eq:3.4c}), in the reference ensemble.
In particular, the Gaussian vertex is determined by
the pair propagator or anomalous
density--density correlation function in the reference ensemble. Denoting the
latter by $C(q)$, the Gaussian term in the LGW functional
reads
\begin{mathletters}
\label{eqs:5.1}
\begin{equation}
\Phi^{(2)}[\Psi] = \sum_{\alpha} \int dq\ {\Psi^{\alpha}}^*(q)\,\bigl[1/\vert 
                   \Gamma^{(c)}\vert - C(q)\bigr]\,\Psi^{\alpha}(q)\quad,
\label{eq:5.1a}
\end{equation}
where we have scaled $\Psi$ with $1/\Gamma^{(c)}$. 
$C(q)$ is the connected propagator
$C(q) = \{\langle n_c(q)\,{\bar n}_c(q)\rangle_0\}_{dis}$, where 
$\{\ldots\}_{dis}$ denotes
the disorder average. $C(q)$ is a complicated correlation
function. However, since the reference ensemble is by construction a Fermi
liquid, the structure of this correlation function is known. Renormalization
group arguments show that the structure of $C$ at low frequencies and long
wavelengths in the limit $T\rightarrow 0$ is\cite{R}
\begin{equation}
C(q) = \frac{Z}{h}\
  \frac{\ln\bigl(\Omega_0/(D{\bf q}^2 + \vert\Omega_n\vert)\bigr)}
     {1 + (\delta k_c/h)\,\ln\bigl(\Omega_0/(D{\bf q}^2
                                             + \vert\Omega_n\vert)\bigr)}\quad,
\label{eq:5.1b}
\end{equation}
Here $\Omega_0 = k_B T_0/\hbar$ is a frequency cutoff on the order of the
Debye frequency (for phonon--mediated superconductivity),
and $\delta k_c$ is the repulsive interaction in the Cooper channel that is
generated within perturbation theory even though the bare $\Gamma^{(c)}$ 
vanishes. $D$ is the diffusion coefficient of the electrons,
$h$ is the renormalized frequency renormalization constant whose
bare value is $H = \pi N_F/8$, see Eq.\ (\ref{eq:3.39a}), 
and $Z$ is the wavefunction renormalization.
All of these parameters characterize the reference ensemble, and it is known
that they provide a complete characterization of $C(q)$\cite{Tc}. Using
Eq.\ (\ref{eq:5.1b}) in Eq.\ (\ref{eq:5.1a}) we obtain, after again
rescaling $\Psi$,
\begin{equation}
\Phi^{(2)}[\Psi] = \sum_{\alpha}\int dq\ {\Psi^{\alpha}}^*(q)\left[t +
   \frac{1}{\ln\bigl(\Omega_0/(D{\bf q}^2 + \vert\Omega_n\vert)\bigr)}\right]
   \Psi^{\alpha}(q)\quad,
\label{eq:5.1c}
\end{equation}
\end{mathletters}%
where $t = -\left(Z\vert K_c\vert - \delta k_c\right)/h$, with
$K_c = \pi N_F^2 \Gamma^{(c)}/4$.

Let us now discuss this result for the Gaussian LGW theory. From the
structure of the Gaussian vertex we immediately read off the values of
the exponents $\eta$, $\gamma$, and $z$, defined as
$\Gamma^{(2)}({\bf q},\Omega = 0) \sim \vert{\bf q}\vert^{2-\eta}$,
$\Gamma^{(2)}({\bf q},\Omega = 0) \sim t^{\gamma}$, and
$\xi_{\tau} \sim \xi^z$, with $\xi_{\tau}$ the relaxation time, see
Eq.\ (\ref{eq:2.1b}). They are
\begin{equation}
\eta = 2\quad,\quad\gamma = 1\quad,\quad z=2\quad,
\label{eq:5.2}
\end{equation}
By scaling
$\vert{\bf q}\vert$ with the correlation length $\xi$,
we also obtain the behavior of the latter,
\begin{equation}
\xi \sim e^{1/2\vert t\vert}\quad.
\label{eq:5.3}
\end{equation}
The exponent $\nu$ therefore does not exist, $\nu = \infty$.

In order to determine the behavior of the order parameter and the
free energy, we need to consider the higher order terms in the
Landau expansion. The coefficient of the quartic term
is a nonlinear anomalous density susceptibility in
the reference ensemble which we denote by $C^{(4)}$.
Due to the cumulant expansion with respect to the
disorder average there are two different contributions to this coefficient,
$C^{(4)} = C_1^{(4)} + C_2^{(4)}$, where $C_1^{(4)}$ is the disorder average
of the four--point correlation function for a given disorder realization,
while $C_2^{(4)}$ is the disorder average of the two--point function squared.
\footnote[24]{The same structure apppears in the ferromagnetic LGW functional.
Here, however, $C_2^{(4)}$ is more important than its analog in the magnetic
case, as we will see below.}
Given that the constant contribution to the Gaussian
coefficient, $C(q)$, is barely the leading term in the limit $q\rightarrow 0$,
and that the quartic coefficient in the magnetic case was singular (see the
discussion after Eq.\ (\ref{eq:4.9})),
one does not expect $C^{(4)}$ to be finite in the limit of vanishing
frequencies and wavenumbers. Indeed, a calculation shows that both $C_1^{(4)}$
and $C_2^{(4)}$ are
singular in this limit. Cutting off the singularity by means of a wave number
$\vert{\bf p}\vert$, one finds for the leading contributions
$C_1^{(4)} \sim u_4/\vert{\bf p}\vert^4\,\ln^4\vert{\bf p}\vert$ and
$C_2^{(4)} \sim v_4/\vert{\bf p}\vert^{4-d}$, respectively, with $u_4$ and
$v_4$ finite coefficicents.
\footnote[25]{This result for $C_2^{(4)}$ holds for $2<d<4$, and we have
 neglected possible logarithmic corrections to the power law.
 For $d>4$, $C_2^{(4)}$ is finite.}
The same method shows that the most divergent contribution
to the coefficient of the term of
order $\vert\Psi\vert^{2n}$ diverges like
\begin{equation}
C^{(2n)} \sim \frac{u_{2n}}{\vert{\bf p}\vert^{4(n-1)}\,\ln^{2n}\vert{\bf p}
                                                             \vert}\quad,
\label{eq:5.4}
\end{equation}
with $u_{2n}$ a finite coefficient.
This implies that the Landau expansion of the cutoff regularized LGW theory
is an expansion in powers of $\Psi/{\bf p}^2\ln(1/\vert{\bf p}\vert)$. The
order parameter field theory, rather than having 
a simple LGW form, is thus again strongly
non--local. This is qualitatively the same effect as in the magnetic
case, but here the singularities one encounters are even stronger.

As in the magnetic case, the
functional $\Phi$ can now be analyzed by using standard
techniques\cite{Ma}. We are looking for a fixed point where the
functional dependence of the 2--point vertex on ${\bf q}$ and $\Omega_n$,
Eq.\ (\ref{eq:5.1c}), is not renormalized. This fixes the exponents $\eta$ and
$z$. As in Secs.\ \ref{sec:II} and \ref{sec:IV}, we 
define the scale dimension of the correlation length to be
$[\xi] = -1$. Power counting then shows that
the coefficients $u_{2n}$ $(n\geq 2)$ of the non--Gaussian terms
have scale dimensions $[u_{2n}]=(n-1)(2-d)$, and hence are irrelevant
operators with respect to the fixed point
for all dimensions $d>2$. $[v_4] = -2(d-2)$ for $2<d<4$,
and the higher cumulants are even more irrelevant. The upper critical dimension
is therefore $d_c^+ = 2$, and for $d>2$ the critical behavior obtained
from the Gaussian theory is exact.

The scaling behavior of the order parameter, and of the free
energy, is determined by the term of $O(\Psi^4)$ which
is a dangerous irrelevant variable with respect
to these quantities. For scaling purposes, the
cutoff wavenumber $\vert{\bf p}\vert$ can be replaced by the inverse
correlation length, $\vert{\bf p}\vert \sim \xi^{-1}$. The scaling
behavior $\Psi \sim {\bf p}^2\ln(1/\vert{\bf p}\vert)$ observed above
then immediately leads to
\begin{equation}
\Psi \sim \frac{\Theta(-t)}{\vert t\vert}\,e^{-1/\vert t\vert}\quad,
\label{eq:5.5}
\end{equation}
One must notice, however,
that the order parameter field $\Psi$ is distinct from the 
physical gap function
$\Delta$. The latter determines the gap in the single--particle excitation
spectrum, and hence scales like the frequency or like $\xi^2$,
\begin{equation}
\Delta \sim \vert t\vert\,\Psi \sim e^{-1/\vert t\vert}\quad.
\label{eq:5.6}
\end{equation}
From Eqs.\ (\ref{eq:5.3}) and (\ref{eq:5.6}) we see that even though the
exponents $\nu$ and $\beta$ do not exist, we can assign a value to their
ratio,
\begin{equation}
\beta/\nu =2 \quad,
\label{eq:5.7}
\end{equation}
in the sense that $\Delta\xi^2 \sim {\rm const.}$

We next consider the critical behavior of the penetration depth $\lambda$ and
the upper critical field $H_{c_2}$. 
Since we have shown that the mean--field/Gaussian
theory yields the exact critical behavior at $T=0$,
all relations between observables that are derived within BCS theory are
valid. In particular, we have $\lambda \sim 1/\sqrt{\Delta}$, and
$H_{c_2} \sim \Delta$ \cite{Tinkham}, which in conjunction with
$\Delta \sim \xi^2$ yields
\begin{equation}
\lambda \sim \xi\quad,\quad H_{c_2} \sim \xi^{-2}\quad.
\label{eq:5.8}
\end{equation}
Similarly, we can determine the scale dimension
of the conductivity $\sigma$ or the resistivity $\rho$. 
The real part of the frequency dependent
conductivity in the superconducting phase has a singular contribution,
which within BCS theory is given by
\begin{equation}
{\rm Re}\,\sigma_s(\Omega) = \frac{\pi^2}{2}\,\sigma_n\,\Delta\,
                                                      \delta (\Omega)\quad,
\label{eq:5.9}
\end{equation}
with $\sigma_n$ the conductivity in the normal state. For scaling purposes,
$\delta(\Omega) \sim 1/\Omega \sim 1/\Delta$. $\sigma_n$ is determined
entirely by properties of the reference ensemble, and hence it does not
show any critical behavior and its scale dimension is zero. We conclude
that the scale dimension of $\sigma_s$ vanishes.
If we assume that the conductivity has only one scaling part, then the
same is true for the conductivity or resistivity in general, and we
obtain
\begin{equation}
\rho(t,T) = \rho(t\ln b,Tb^2)\quad.
\label{eq:5.10}
\end{equation}
The resistivity thus shows a step discontinuity from a finite value to
zero as one crosses the phase boundary at $T=0$.

We now turn to the free energy density $f$. Hyperscaling suggests that
$f$ scales like $f \sim T/V \sim \xi^{-(d+2)}$, which leads to a
homogeneity law
\begin{equation}
f(t,T,u_4) = b^{-(d+2)}\,f(t\ln b, T b^2, u_4 b^{2-d})\quad,
\label{eq:5.11}
\end{equation}
where of the irrelevant operators we have written only $u_4$ explicitly.
$f$ is proportional to $u_4\,\Delta^4 \sim 1/u_4$, and hence the
effective scale dimension of $f$ is $[f] = 4$, which yields
\begin{equation}
f(t,T) = b^{-4}\,f(t\ln b,Tb^2)\quad,
\label{eq:5.12}
\end{equation}
Hyperscaling is violated by the usual mechanism that is operative above
an upper critical dimension, viz. by means of the quartic coefficient
being dangerously irrelevant with respect to the free energy. The more
exotic violation mechanisms that we encountered in the magnetic case
are not realized here. By
differentiating twice with respect to $T$ one obtains the specific heat
coefficient $\gamma(t,T)$,
\begin{equation}
\gamma(t,T) = \gamma(t\ln b,Tb^2)\quad.
\label{eq:5.13}
\end{equation}
This implies, among other things, that the specific heat coefficient
approaches a constant as the temperature is lowered to zero at the
critical coupling strength. More detailed calculations show a step
discontinuity in $\gamma(T=0)$ at $t=0$.

The quartic term whose coefficient is $C_2^{(4)}$
yields corrections to scaling. $C_2^{(4)}$ represents fluctuations
in the position of the critical point: By making the coupling
constant $\Gamma^{(c)}$ a random variable, 
and integrating out that randomness, one
obtains a term of that structure. By repeating the arguments of
Ref.\ \onlinecite{WeinribHalperin}, one finds that the relative fluctuations
of the position of the critical point decay anomalously slowly, viz.
$\Delta t/t \sim \xi^{-(d-2)}$. Translating that into the corresponding
fluctuations of $T_c$ via $T_c = T_0\,\exp(-1/\vert t\vert)$ we obtain
\begin{equation}
\Delta T_c/T_c \sim T_c^{(d-2)/2}\,F\bigl(\ln (T_0/T_c)\bigr)\quad,
\label{eq:5.14}
\end{equation}
with $F(x)$ a function that depends parametrically on the disorder.
Since it depends on $T_c$ only logarithmically, the latter dependence is
weak. Equation (\ref{eq:5.14}) predicts very strong disorder fluctuations
in thin superconducting films. This prediction is in semi--quantitative
agreement with recent experiments by Hsu et al.\cite{HsuChervenakValles}.

\section{Metal--Insulator Transitions}
\label{sec:VI}

\subsection{Disordered Fermi liquids}
\label{subsec:VI.A}

In recent years there has been a considerable amount of work on using 
renormalization group
ideas to derive and justify Landau's Fermi liquid theory\cite{FLT}. The main
motivation was the idea that non--Fermi liquid notions might be important for
understanding, for example, the normal state of the high $T_c$ materials,
and that in order to derive and discuss non--Fermi liquid behavior, 
one must first have a
deeper understanding of why most systems are Fermi liquids. In
this context, renormalization group 
ideas are used to understand an entire phase and not just a
single (although important) critical point 
in the phase diagram. Technically this
implies that one must look for completely stable renormalization group 
fixed points, as opposed
to the more usual critical fixed points that are associated with phase 
transitions.

In this subsection we briefly review how renormalization group 
ideas can be applied to derive
and justify the existence of a disordered Fermi liquid phase for $d>2$, and,
in addition, can be used to derive what are commonly known as weak
localization effects\cite{fermions}.
\footnotemark[20]

\subsubsection{The disordered Fermi liquid fixed point}
\label{subsubsec:VI.A.1}

To proceed we parametrize the $\hat{Q}$ in Eq. (\ref{eq:3.39b}) 
in analogy with the
parameterization of $\hat{\phi}$ given by Eqs. (\ref{eqs:3.23}). 
We write $\hat{Q}$ in a block matrix form as,

\begin{equation}
\hat{Q}=\left( 
           \matrix{\sqrt{1-qq^{\dagger}} & q \cr
                 q^{\dagger}  & -\sqrt{1-q^{\dagger}q}\cr}
                                                 \right) \quad,
\label{eq:6.1}
\end{equation}
where the matrix $q$ has elements $q_{nm}$ whose frequency labels
are restricted to $n\geq 0,m<0.$ Now, while the sigma--model, 
Eq. (\ref{eq:3.39a}),
can be expressed entirely in terms of $q$, the corrections to the sigma--model
action also depend on
\begin{equation}
\Delta P = P - \langle P\rangle\quad. 
\label{eq:6.2}
\end{equation}
Considering the total action,
\begin{equation}
{\cal A} ={\cal A}_{NL\sigma M} + \Delta {\cal A}\quad, 
\label{eq:6.3}
\end{equation}
we perform a momentum shell renormalization group procedure. 
For the rescaling part
of this transformation, we need to assign scale dimensions to the soft field 
$q$, and to the massive field $\Delta P$ as well. If the scale dimension of
a length $L$ is $[L] = -1$, we write, in analogy to scaling near a critical
point,
\begin{mathletters}
\label{eqs:6.4}
\begin{equation}
[q({\bf x})] = \frac{1}{2}\,(d-2+\eta^{\prime})\quad, 
\label{eq:6.4a}
\end{equation}
\begin{equation}
[\Delta P({\bf x})] = \frac{1}{2}\,(d-2+\eta)\quad,
\label{eq:6.4b}
\end{equation}
\end{mathletters}%
which defines the exponents $\eta$ and $\eta^{\prime}$. The
stable Fermi--liquid fixed point of the theory is characterized by the choice
\begin{mathletters}
\label{eqs:6.5}
\begin{equation}
\eta =2\quad,\quad\eta^{\prime} = 0\quad. 
\label{eq:6.5a}
\end{equation}
Physically, $\eta^{\prime} = 0$ corresponds to diffusive
correlations of $q$, and $\eta = 2$ means that the correlations of the
$\Delta P$ are of short range. This is indeed what one expects in a
disordered Fermi liquid. In addition, we must specify the scale dimension of
frequency and temperature, i.e., the dynamical scaling exponent 
$z=[\omega]=[T]$. In order for the fixed point to be consistent 
with diffusion, that is with
frequencies that scale like the squares of wavenumbers, we must choose
\begin{equation}
z=2\quad. 
\label{eq:6.5b}
\end{equation}
\end{mathletters}%

Now we expand the sigma--model action, Eq. (\ref{eq:3.39a}), 
in powers of $q$. In a
symbolic notation that leaves out everything not needed for power counting
purposes, we write,
\begin{equation}
{\cal A}_{NL\sigma M} = \frac{1}{G}\int d{\bf x}\,(\nabla q)^2 
                         + H\int d{\bf x}\,\omega\,q^2
   + \Gamma T\int d{\bf x}\,q^2 + O(\nabla^2 q^4,\omega q^4,Tq^3)\quad,
\label{eq:6.6}
\end{equation}
with the bare coupling constants $G\sim 1/\sigma_0$ and $H\sim N_F$. 
$\Gamma$ can stand for any of the three interaction constants
$\Gamma^{(s)}$, $\Gamma^{(t)}$, or $\Gamma^{(c)}$. Power
counting gives
\begin{equation}
[G]=[H]=[\Gamma ] = 0\quad.
\label{eq:6.7}
\end{equation}
These terms therefore make up part of the fixed point action.

Next consider the corrections that arise within the sigma--model. The leading
ones are indicated in Eq.\ (\ref{eq:6.6}). We denote the corresponding coupling
constants by $C_{\nabla ^2q^4}$, etc., with a subscript that identifies the
structure of the respective contribution to the action. One finds,
\begin{equation}
[C_{\nabla ^2q^4}] = [C_{\omega q^4}^{}] = [(C_{T q^3})^2] = -(d-2)\quad.
\label{eq:6.8}
\end{equation}
In the last equality in Eq.\ (\ref{eq:6.8}) we have considered the square of
$C_{T q^3}$, since any contributions to physical correlation functions
contain the corresponding term squared. We see that all of these operators 
are irrelevant with respect
to the disordered Fermi--liquid fixed point for all $d>2$, and that they become
marginal in $d=2$ and relevant in $d<2$. All other terms in the sigma--model
are at least as irrelevant as those considered above.

It is easy to show that $\Delta {\cal A}$ 
in Eq. (\ref{eq:6.3}) also contributes to the disordered
Fermi--liquid fixed point action. We denote this contribution by 
$\Delta {\cal A}^{*}$. An inspection shows that
$\Delta {\cal A}^{*}$ depends only
on $\Delta P$ and does not couple to $q$. Further, all of the corrections to
$\Delta {\cal A}$ are more irrelevant for $2<d<4$ than the operators with
scale dimensions equal to 
$2-d$ that are indicated in Eq. (\ref{eq:6.8}). 
The scale dimension of the least irrelevant additional corrections is equal
to $-2$.

The above renormalization group arguments show that the theory contains 
a disordered
Fermi--liquid fixed point that is stable for all $d>2$. 
The effective fixed point action is,
\begin{equation}
{\cal A}_{eff}^{*} = {\cal A}_{NL\sigma M}[q] + \Delta {\cal A}^{*}
                                                     [\Delta P]\quad,
\label{eq:6.9}
\end{equation}
and the leading irrelevant terms near this fixed point have scale
dimensions given by Eq.\ (\ref{eq:6.8}).

\subsubsection{Scaling behavior of observables}
\label{subsubsec:VI.A.2}

We now discuss the physical meaning of the corrections to scaling induced by
the irrelevant operators that we have identified above. Let us denote by the
generic name $u$ any of the least irrelevant operators whose scale dimension
is $[u]=-(d-2)$, and let us discuss various observables, viz. the
conductivity $\sigma$, the specific heat coefficient $\gamma_V$, the
single--particle density of states $N$, and the spin susceptibility $\chi_s$.
Which of the various operators with scale dimension $-(d-2)$ is the important
one depends on the quantity under consideration.

Let us first consider the dynamical conductivity, $\sigma (\omega)$. Its
bare value is proportional to $1/G$, and according to Eq. (\ref{eq:6.7}) 
its scale dimension is zero. We therefore have the scaling law,
\begin{mathletters}
\label{eqs:6.10}
\begin{equation}
\sigma (\omega ,u)=\sigma (\omega b^z,ub^{-(d-2)})\quad,
\label{eq:6.10a}
\end{equation}
where $b$ is an arbitrary renormalization group scale factor. 
By putting $b=1/\omega^{1/z}$, and using $z=2$, Eq. (\ref{eq:6.5b}), 
as well as the fact that $\sigma (1,x)$
is an analytic function of $x$, we find that the conductivity has a
singularity at zero frequency, or a long--time tail, of the form
\begin{equation}
\sigma (\omega) \sim {\rm const} + \omega^{(d-2)/2}\quad.
\label{eq:6.10b}
\end{equation}
\end{mathletters}%
This nonanalyticity is well known from perturbation theory for
both noninteracting and interacting electrons\cite{AAKL,LeeRamakrishnan}. 
The above analysis proves
that the $\omega ^{(d-2)/2}$ is the exact leading nonanalytic behavior.

The specific heat coefficient, $\gamma_V = c_V/T$, is proportional to the
quasiparticle density of states 
$H$\cite{CastellaniDiCastro,CastellaniKotliarLee,Castellanietal},
whose scale dimension vanishes
according to Eq.\ (\ref{eq:6.7}). We thus have a scaling law
\begin{mathletters}
\label{eqs:6.11}
\begin{equation}
\gamma_V(T,u) = \gamma_V(Tb^z,ub^{-(d-2)})\quad,
\label{eq:6.11a}
\end{equation}
which leads to a low-temperature behavior
\begin{equation}
\gamma_V(T) \sim {\rm const} + T^{(d-2)/2}\quad.
\label{eq:6.11b}
\end{equation}
\end{mathletters}%
From perturbation theory it is known\cite{AAKL} that $\gamma_V$ shows this
behavior only for interacting electrons, while for non--interacting systems
the prefactor of the nonanalyticity vanishes. This can not be seen by our
simple counting arguments.

The single--particle density of states, $N$, is proportional to the
expectation value of $Q$, and to study the leading correction to the finite
fixed point valueof $N$ it suffices to replace $Q$ by $\hat{Q}$. 
Then we have, in
symbolic notation, $N\sim 1 + \langle qq^{\dagger}\rangle+\ldots = 1+\Delta N$. The scale
dimension of $\Delta N$ is $[\Delta N] = 2[q] = d-2$. We find the scaling law
\begin{mathletters}
\label{eqs:6.12}
\begin{equation}
\Delta N(\omega) = b^{-(d-2)}\,\Delta N(\omega b^z)\quad,
\label{eq:6.12a}
\end{equation}
which leads to the so-called Coulomb anomaly\cite{AltshulerAronov},
\begin{equation}
N(\omega) \sim {\rm const} + \omega^{(d-2)/2}\quad. 
\label{eq:6.12b}
\end{equation}
\end{mathletters}%
Again, this behavior is known to occur only in the presence of
electron-electron interactions.

Finally, we consider the static, wave vector dependent spin susceptibility, 
$\chi_0({\bf q})$. 
$\chi_0$ is given by a $Q$--$Q$ correlation function, and the leading
correction to the finite Fermi--liquid value is obtained by replacing both of
the $Q$ by $q$. Then we have a term of the structure 
$\chi_0\sim T\int d{\bf x}\ q^{\dagger}q$, with scale dimension $[\chi_0]=0$. 
The relevant scaling law is
\begin{mathletters}
\label{eqs:6.13}
\begin{equation}
\chi_0({\bf q},u) = \chi_0({\bf q},ub^{-(d-2)})\quad,
\label{eq:6.13a}
\end{equation}
which leads to a nonanalytic dependence on the wave number,
\begin{equation}
\chi_0({\bf q}) \sim {\rm const} - \vert{\bf q}\vert^{(d-2)}\quad.
\label{eq:6.13b}
\end{equation}
\end{mathletters}%
This behavior is also known from perturbation theory\cite{fmdirty}, and holds
only for interacting electrons. As we have seen in Sec.\ \ref{subsubsec:IV.A.1}
above, it has interesting consequences for the theory of ferromagnetism.

To summarize, we see from the above arguments that all of the so--called
weak--localization effects, i.e. nonanalytic dependencies of various
observables on frequency, temperature, or wave number, in disordered electron
systems that are well known from perturbation theory, emerge naturally in
the present context as the leading corrections to scaling near the
Fermi--liquid fixed point of a general field theory for disordered interacting
electrons. Apart from providing an aesthetic, unifying, and very simple
explanation for these effects, our arguments also prove that they do indeed
constitute the leading nonanalytic behavior, a conclusion that cannot be
drawn from perturbation theory alone. 

We finally note that nonanalyticities that are very similar to those
discussed above occur in classical fluids. In that context
they are known as long--time tail effects, and they were first discussed
theoretically by using many--body perturbation theory and mode coupling
theory\cite{DorfmanTRKSengers,ernst}. Later, they were examined using 
renormalization group ideas, and they were shown to be
related to corrections to the scaling behavior near a hydrodynamic fixed 
point\cite{ForsterNelsonStephen}.

\subsection{The Anderson--Mott transition}
\label{subsec:VI.B}

It is well known that at sufficiently large disorder, the metallic disordered
Fermi liquid phase discussed in the last subsection becomes unstable
against an insulating phase. Such metal--insulator transitions (MITs)
are observed in doped semiconductors and other disordered electron systems, 
and the generalized nonlinear sigma--model shown in
Eq.\ (\ref{eq:3.39a}) is capable of describing them\cite{R}. 
Metal--insulator transitions whose critical
behavior is determined by both the electron--electron interaction and the
disorder, are commonly referred to as Anderson--Mott transitions, to
distinguish them from purely disorder driven MITs (`Anderson transitions')
and purely correlation driven ones (`Mott transitions'), respectively.
They can be grouped into
two broad classes: (1) Those that are related to fixed points in the
vicinity of $d=2$, and (2) one related to a Gaussian fixed point in high
($d>6$) dimensions. Somewhat strangely, these transitions seem to be quite
different in nature, and attempts to extrapolate down from $d=6$ to $d=3$
yields results that are incompatible with attempts to extapolate up from
$d=2$ to $d=3$. As a result, the theoretical description of the MITs that
are observed in $3$--$d$ systems is actually an open problem, the substantial
amount of effort that has gone into this subject
notwithstanding. We therefore do not
pretend here to be able to give a coherent theoretical picture. Rather, we
present the results near $d=2$ and the results in high dimensions separately,
and then add some speculations about the behavior in $d=3$, with emphasis
on the points that are not understood.

\subsubsection{Anderson--Mott transition near two dimensions}
\label{subsubsec:VI.B.1}

This subject has been reviewed in Ref.\ \onlinecite{R}, and we
refer the reader there for details and references to the original work.
Here we give a brief, updated overview designed to tie this subject in with the
more recent developments that are reviewed elsewhere in this chapter.

The MITs near $d=2$ fall into several distinct universality classes that
are related to external fields that couple to various degrees of freedom
in the spin quaternion space defined in Sec.\ \ref{subsubsec:III.C.2}. 
Specifically,
an external magnetic field (to be referred to as MF), magnetic impurities
(MI), and spin--orbit scattering (SO) all break some of the additional
symmetries mentioned after Eq.\ (\ref{eq:3.33}) that lead to $Q$--matrix
sectors $^i_rQ$ with $i,r\neq 0$ being soft 
modes\cite{EfetovLarkinKhmelnitskii}. The $r=0,i=0$ sector
is directly controlled by the Ward identity, Eq.\ (\ref{eq:3.33}), and
it always is soft. The universality class with no additional symmetry
breakers is called the generic class, and is denoted by G.
Table\ \ref{table:6.1} lists the remaining soft
modes for all of these universality classes. 
\mediumtext
\begin{table}[thb]
\caption{Universality classes and soft modes for the MIT near $d=2$.}
\label{table:6.1}
\medskip
\begin{tabular}{ccc}
Universality class   &   Symmetry breaker       &   Soft modes\\
\tableline
\\
MF                   &   magnetic field         &   $r=0,3\ ;\ i=0,3$\\
\\
MI                   &   magnetic impurities    &   $r=0,3\ ;\ i=0$\\
\\
SO                   &   spin--orbit scattering &   $r=0,1,2,3\ ;\ i=0$\\
\\
G                    &   none                   &      all\\
\\
\end{tabular}
\end{table}
\widetext
\begin{table}[thb]
\caption{Values for the three independent exponents $\nu$, $\beta$, and $z$
 for the eight universality classes in $d=2+\epsilon$. Values are given for
 $d=2+\epsilon$ dimensions except for class G, where approximate values
 for $d=3$ based on a two--loop approximation in
 Ref.\ \protect\onlinecite{Guniversalityclass} are shown, and for class SO (SR),
 where the value shown for $\nu$ is the numerical result given in
 Ref.\ \protect\onlinecite{SOSRuniversalityclass}. $\beta$ for the class MF (SR)
 is known to one-loop order, but depends on non--universal quantities, see
 Ref.\ \protect\onlinecite{R}. The class G (SR) has never been considered.}
\medskip
\begin{tabular}{ccccccccc}
 &\multicolumn{2}{c}{MI}&\multicolumn{2}{c}{MF}&\multicolumn{2}{c}{SO}
 &\multicolumn{2}{c}{G}\\
 & SR & LR     &    SR & LR     &    SR & LR     &     SR & LR\\
Exponent\\
\tableline
\\
$\nu$   & $\frac{1}{2\epsilon} - \frac{3\epsilon}{4} + O(\epsilon^3)$
        & $\frac{1}{\epsilon} + O(1)$
        & $\frac{1}{\epsilon} + O(1)$
        & $\frac{1}{\epsilon} + O(1)$
        & $\approx 1.3\pm 0.2$
        & $\frac{1}{\epsilon} + O(1)$
        & ?
        & $\approx 0.75$\\
\\
$\beta$ & 0
        & $\frac{1}{\epsilon} + O(1)$
        & non--universal
        & $\frac{1/2\epsilon}{1-\ln 2} + O(1)$
        & $0$
        & $\frac{2}{\epsilon} + O(1)$
        & ?
        & $\approx 0.50$\\
\\
$z$     & $d$
        & $2 + \frac{\epsilon}{2} + O(\epsilon^2)$
        & $d$
        & $d$
        & $d$
        & $2 + O(\epsilon^2)$
        & ?
        & $\approx 5.91$\\
\\
\end{tabular}
\label{table:6.2}
\end{table}
In addition, the critical
properties turn out to depend on the type of electron--electron interaction
considered; they are different depending on whether a short--ranged model
interaction is used, or a long--range Coulomb interaction.
\footnote[26]{The short--ranged case is not entirely academic. 
It has been proposed to realize it
by putting a grounded metallic plate behind the sample which cuts off the
long--ranged Coulomb interaction by means of the image charges induced in the
plate\cite{ShortRangeExperiment}.}
For each of the four universality classes G, MF, MI, and SO one therefore
distinguishes between short--ranged (SR) and long--ranged (LR) subclasses,
leading to a total of eight distinct universality classes. For the three
independent critical exponents (cf. Sec.\ \ref{sec:II}) we choose the
correlation length exponent $\nu$, the dynamical exponent $z$, and the
exponent $\beta$ that describes the vanishing of the density of states
at the Fermi surface, $N(\epsilon_F)\sim t^{\beta}$. Their values for
the various universality classes are shown in Table\ \ref{table:6.2}.
Other critical exponents of interest are related to these three by means
of exponent relations that are listed in Table\ \ref{table:6.3}. Which
quantities are critical and which are not depends on which modes are soft,
and therefore on the universality class. This is also indicated in
Table\ \ref{table:6.3}.
\begin{table}[htb]
\caption{Physical quantities, their critical behavior, and relevant exponent
 relations. Also listed are the universality classes for which the respective
 quantities show critical behavior. The order parameter density is the density
 of states. The first two columns are generally valid, while the last two
 pertain to the Anderson--Mott transition near $d=2$ only. In high dimensions,
 there is only one universality class, and some exponent relations change,
 see Sec.\ \protect\ref{subsubsec:VI.B.3}.}
\medskip
\begin{tabular}{cccc}
Physical quantity & Scaling behavior & Exponent relation & Universality classes
\\
\tableline
\\
Correlation length
                   & $\xi(t)\sim t^{-\nu}$
                   & independent exponent
                   & all\\
\\
Correlation time
                   & $\xi_{\tau}\sim \xi^z$
                   & independent exponent
                   & all\\
\\
Density of states
                   & $N(\epsilon_F,t)\sim t^{\beta}$
                   & independent exponent
                   & MI(LR), MF, SO(LR), G\\
\\
Electrical conductivity
                   & $\sigma(t,T=\omega=0)\sim t^s$
                   & $ s=\nu (d-2)$
                   & all\\
\\
Specific heat
                   & $c_V(t=0,T) \sim T^{1-\kappa/z}$
                   & $\kappa = z-d$
                   & MI(LR), SO(LR)\\
\\
Heat diffusion coefficient
                   & $D_h(t,T=0) \sim t^{s_h}$
                   & $s_h = \nu (z-2)$
                   & MI(LR), SO(LR), G\\
\\
Spin diffusion coefficient
                   & $D_s(t,T=0) \sim t^{s_s}$
                   & $s_s = \nu (z-2)$
                   & G\\
\\
Order parameter density
                   &$G({\bf k},t=T=0)\sim\vert{\bf k}\vert^{-2+\eta}$
                   & $\eta = 2-d+2\beta/\nu$
                   & MI(LR), MF, SO(LR), G\\
correlation function\\
\\
Order parameter density
                   & $\chi_{OP}(t,T=0)\sim t^{-\gamma}$
                   & $\gamma = \nu d - 2\beta$
                   & MI(LR), MF, SO(LR), G\\
susceptibility\\
\\
\end{tabular}
\label{table:6.3}
\end{table}

The theoretical analysis of these universality classes near $d=2$
starts from the nonlinear sigma--model, Eq.\ (\ref{eq:3.39a}), and
the parametrization, Eq.\ (\ref{eq:6.1}), of the $Q$--matrix or an
equivalent parametrization. This procedure does {\em not} make use
of the order parameter field theory derived in Sec.\ \ref{subsubsec:III.C.4},
and the fact that the density of states is the order parameter for the
transition does not become apparent. Rather, one 
constructs a loop expansion that
is equivalent to an expansion in powers of $q$, and proceeds with a
renormalization group analysis analogous to that for the $O(N)$--Heisenberg
model in $d=2+\epsilon$ dimension\cite{ONSigmaModel}. 
For the universality classes
MI, MF, and SO, this procedure readily leads to fixed points corresponding
to MITs, and the asymptotic critical properties at these transitions
have been worked out to lowest order in a $2+\epsilon$ 
expansion\cite{Finkelstein2}.
For the conductivity and the density of states, 
the results are in reasonable agreement with
experiments, although the ill--behaved nature of the $2+\epsilon$ expansion
precludes quantitative comparisons\cite{R}. Measurements of the
spin susceptibility and the specific heat, on the other hand, have
shown non--Fermi liquid like behavior on either side of the 
MIT\cite{thermodynamics}, which the
nonlinear sigma--model can not explain. Very similar results were obtained
experimentally on systems that are believed to be in the universality class
G\cite{thermodynamicsG}. These features are usually
interpreted in terms of local magnetic moments that are believed to be
ubiquitous in disordered electronic systems\cite{SachdevRoySoc}. 
Currently no unified theory
exists that can describe the interplay between these local moments and
the critical behavior of the transport properties at an MIT.

For the universality classes MI (SR) and SO (SR) it turns out that the
electron--electron interaction amplitude scales to zero under 
renormalization\cite{CDLM}.
The critical behavior in these cases is therefore the same as for the
corresponding universality classes
of non--interacting electrons\cite{LeeRamakrishnan,MacKinnonKramer}.
The density of states is uncritical ($\beta=0$), and the dynamical
exponent is $z=d$. For the class MI (SR) one finds a fixed point at one--loop
order that corresponds to a MIT, with a correlation length exponent
$\nu$ as given in Table\ \ref{table:6.2}. The
behavior for the class SO (SR) was unclear for a long time, since in this
case there is no fixed point that is accessible by means of a 
$2+\epsilon$ expansion. However, there is now good evidence from both numerical
analyses and high--order perturbation theory for a MIT in 
$d=3$\cite{Wegner89,Hikami,SOSRuniversalityclass}.

For the universality classes MI (LR), SO (LR), and MF, the renormalization
group analysis yields MIT fixed points where the interaction strength scales
to a finite number, so that the critical behavior is different from that of
noninteracting electrons. The results are listed in Table\ \ref{table:6.2}.
For the MI and MF universality classes the Cooper channel ($r=1,2$ in the
spin--quaternion basis) does not contribute to the soft modes, see
Table\ \ref{table:6.1}. It is therefore sufficient to renormalize the
particle--hole degrees of freedom. For the class SO (LR), on the other
hand, one has to deal with the Cooper channel as well. How to properly
do this in an interacting system has given rise to some controversy. 
The current state of affairs
is that no satisfactory renormalization group treatment of the Cooper
channel interaction amplitude $\Gamma^{(c)}$ exists\cite{cooperons}. 
However, there is general agreement that $\Gamma^{(c)}$ is irrelevant at
the MIT fixed point, so that the asymptotic critical behavior is unaffected
by this unsolved problem. It is also likely that $\Gamma^{(c)}$ is only
marginally irrelevant, and gives rise to logarithmic corrections to
scaling\cite{logs,cooperons}.

The behavior one encounters in class G is substantially more complicated
than that in any of the other universality classes. Low--order applications
of the perturbative renormalization group reveal a runaway behavior of the
spin--triplet interaction constant, $\Gamma^{(t)}$, and the flow equations
do not allow for a fixed point corresponding to a MIT\cite{FinkelsteinZPhys}. 
There is a popular
belief that this runaway flow is an indication of local--moment formation,
which weak--disorder perturbation theory is inadequate to 
describe\cite{SachdevRoySoc}. We do not believe that this is the correct
interpretation of the runaway flow, for the following reasons. As has
been shown in Ref.\ \onlinecite{IFS}, the leading singularities in the
runaway region, i.e. for the renormalized $\Gamma^{(t)}\rightarrow\infty$,
can be resummed exactly to all orders in the loop expansion. The result
is a pair of coupled integral equations for the spin and heat diffusion
coefficients, the solution of which yields a phase transition where the
spin diffusion coefficient vanishes. This transition has also been
studied with renormalization group techniques, which corroborated the
results obtained from the integral equations\cite{IFS2}. The conductivity
or charge diffusion coefficient was found to be uncritical, and decoupled 
from the spin transport in the critical region, so the transition is not
a MIT. A very remarkable aspect of these results was that the critical
behavior could be obtained exactly, apart from possible logarithmic
corrections, in all dimensions $d>2$. The reasons for this unusual
feature were not clear at the time.
Originally, the nature of this transition was not known, and the
original papers speculated about it leading to a pseudomagnetic phase
with `incompletely frozen spins'. While such an `incompletely frozen spin
phase' might have had something to do with local moments, later developments
showed that these speculations were incorrect. The exact critical behavior found
in Refs.\ \onlinecite{IFS,IFS2} turns out to be identical with that of
the exactly soluble paramagnet--to--ferromagnet transition in disordered
itinerant electron systems that was reviewed in Sec.\ \ref{subsubsec:IV.A.1}
above. This can hardly be a coincidence, and we therefore believe that the
correct interpretation of the runaway flow found in the G universality
class is a transition to a long--range ordered ferromagnetic phase, not
an instability against local moment formation. The fact that the runaway
flow occurs for arbitrarily small values of the disorder and the
interaction strength, provided that one works close to $d=2$, suggests
a phase diagram for low--dimensional systems as shown in Fig.\ \ref{fig:6.1}.
\begin{figure}[t]
\centerline{\psfig{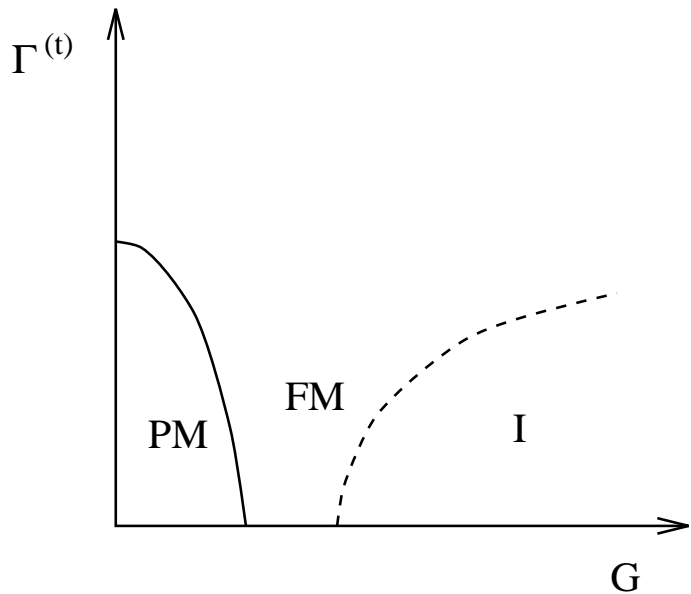}\vspace*{5mm}}
\caption{Schematic phase diagram for disordered itinerant electron systems
at $T=0$ close to $d=2$, in the plane spanned by the spin--triplet
interaction constant $\Gamma^{(t)}$
and the disorder $G$. The phases shown are the paramagnetic
metal (PM), the ferromagnetic metal (FM), and the insulator (I). It is not
known whether there is another phase transition within I from a ferromagnetic
to a paramagnetic insulator.}
\label{fig:6.1}
\bigskip
\end{figure}
The point is that close to two dimensions, the paramagnetic metal phase and
the insulator phase are always separated by a ferromagnetic phase. This is
the reason why the MIT in the G universality class is not accessible by
means of an $\epsilon$--expansion about $d=2$, and it is what is causing the
runaway flow in the $2+\epsilon$ expansion. Whether or not the
paramagnetic metal phase is unstable against the formation of local moments,
and if so, why this physical feature is not reflected in the sigma--model,
is a separate question. We also point out that there are other competing
instabilities in this universality class, in addition to the ones noted
above. For example, in Ref.\ \onlinecite{triplet_sc} we have argued that in 
$d=2$, there is an instability
to a novel type of even--parity superconductivity. All of this suggest that
the actual phase diagram in and near two dimensions is very complicated and,
at this point, not understood.

For $d>2+\epsilon_c$, with $\epsilon_c$ a critical value that was estimated
in Ref.\ \onlinecite{IFS2}, there is a direct transition from the 
paramagnetic metal to an insulating phase, see Fig.\ \ref{fig:6.2}.
\begin{figure}[t]
\centerline{\psfig{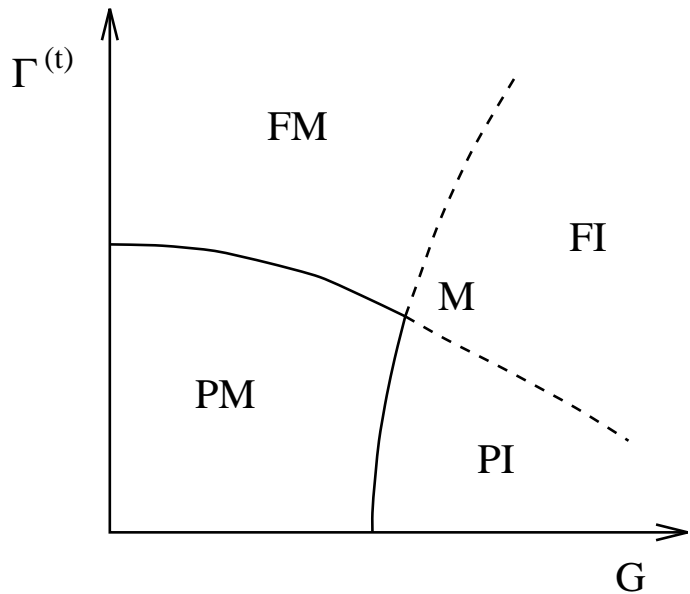}\vspace*{5mm}}
\caption{Schematic phase diagram for a $3$--$d$ disordered itinerant electron
system in the plane spanned by the spin--triplet interaction constant
$\Gamma^{(t)}$ and the disorder $G$. The phases shown are a paramagnetic
metal (PM), a ferromagnetic metal (FM), a paramagnetic insulator (PI), and
a ferromagnetic insulator (FI). M denotes a multicritical point.}
\label{fig:6.2}
\bigskip
\end{figure}
Obviously, this transition cannot be described by means of a controlled
$\epsilon$--expansion about $d=2$. The properties of this transition in
$d=3$ have been estimated by means of a two--loop 
approximation\cite{Guniversalityclass}, the
results of which are shown in Table\ \ref{table:6.2}.

Finally, we mention that all existing treatments of the Anderson--Mott
transition conclude that $d_c^- = 2$ is the lower critical dimension of
the problem, in the sense that in $d=2$ there is no metallic phase for
any degree of disorder. Until recently, the experimental situation
was in agreement with this. Experiments that seem to show a transition
from an insulating phase to a conducting one in $2$--$d$ Si MOSFET
systems therefore came as a considerable surprise\cite{MOSFETexperiment}.
Several theoretical speculations have been put forward\cite{MOSFETtheory},
but the nature of this phenomenon is currently unclear.

\subsubsection{Anderson--Mott transition in high dimensions}
\label{subsubsec:VI.B.2}

Let us now return to the generalized sigma--model in the form given by
Eq.\ (\ref{eq:3.42}), where the soft modes have been integrated out.
This rewriting of the action one would assume to lead to a simple
formulation of the problem in sufficiently high dimensions, where the
soft modes do not provide the leading physical effects. This expectation
will prove to be correct, and will allow us to determine the exact critical
behavior at a metal--insulator transition in dimensions $d>6$\cite{opmit}.

We look for saddle--point solutions $Q_{sp}$, $\Lambda_{sp}$
that are spatially uniform and satisfy
\begin{mathletters}
\label{eqs:6.14}
\begin{equation}
{_{r}^{i}(Q_{sp})_{nm}^{\alpha\beta}} =
\delta_{r0}\delta_{i0}\delta_{nm}\delta_{\alpha\beta}N_{n}^{(0)}\quad,
\label{eq:6.14a}
\end{equation}
\begin{equation}
{_{r}^{i}(\Lambda_{sp})_{nm}^{\alpha\beta}} =
\delta_{r0}\delta_{i0}\delta_{nm}\delta_{\alpha\beta}\ell_{n}^{(0)}\quad,
\label{eq:6.14b}
\end{equation}
\end{mathletters}%
where the superscript $(0)$ denotes the saddle--point approximation.
The replica, frequency, and spin--quaternion structures in
Eqs.\ (\ref{eqs:6.14}) are motivated by the fact
that $\langle {^{i}_{r}Q_{nm}^{ij}}\rangle$ and
$\langle {^{i}_{r}\Lambda_{nm}^{ij}}\rangle$ have these properties,
and that in the mean--field approximation averages are
replaced by the corresponding saddle--point values.

Taking the extremum of the action, Eq.\ (\ref{eq:3.42}), with respect to
$\Lambda$ or $\ell^{(0)}$ one obtains
\begin{mathletters}
\label{eqs:6.15}
\begin{equation}
\bigl(N_{n}^{(0)}\bigr)^{2} = 1 - G\,\Gamma^{(s)}\,f_{n}(\ell^{(0)})\quad,
\label{eq:6.15a}
\end {equation}
and taking the extremum with respect to $Q$ or $N^{(0)}$ gives
\begin{equation}
\ell_{n}^{(0)} = 2GH\omega_{n}/N_{n}^{(0)}\quad.
\label{eq:6.15b}
\end{equation}
\end{mathletters}%
$f_n$ in Eq.\ (\ref{eq:6.15a}) is a functional of $\ell$ that contains
several frequency integrations over $\ell^{(0)}_m$, and that in addition
depends parametrically on the external frequency $\omega_n$, on the
disorder, and on the interaction constants\cite{opmit}. For simplicity we show
only the spin--singlet interaction, $\Gamma^{(s)}$, explicitly.
It has been shown in Ref.\ \onlinecite{opmit} that the other interaction
channels do not qualitatively change the conclusions.

$N_n$ is simply related to the density of states through Eq.\ (\ref{eq:3.31}).
For fixed $\Gamma^{(s)}$, Eqs.\ (\ref{eqs:6.15}) describe a decrease
of the density of states with increasing disorder $G$.
Furthermore, $N(\epsilon_F + \omega)$
is nonanalytic at $\omega=0$. An iteration to first order in
$G$ recovers the `Coulomb anomaly' that is well known from perturbation
theory\cite{AltshulerAronov}, see Eq.\ (\ref{eq:6.12b}) above.
With further increasing disorder, $N(\epsilon_F)$
vanishes at a critical value $G_c$ of the disorder.
The mean--field approximation thus describes a phase
transition with a vanishing density of states at the Fermi level, 
the hallmark of the 
Anderson--Mott transition.
(Transport properties we will discuss shortly). Note that in the absence of
interactions,
$N^{(0)}\equiv 1$. This reflects the fact that the density of states 
does {\it not}
vanish at an Anderson transition\cite{WegnerDOS}.

The saddle point equations, Eqs.\ (\ref{eqs:6.15}), together with an
expansion to Gaussian order about the saddle point, determine the critical
behavior of this Anderson--Mott transition in high dimensions. One finds
standard mean--field/Gaussian values for all static exponents,
\begin{mathletters}
\label{eqs:6.16}
\begin{equation}
\beta = \nu = 1/2\quad,\quad \gamma = 1\quad,\quad \eta = 0\quad,\quad
\delta = 3\quad,
\label{eq:6.16a}
\end{equation}
and for the dynamical critical exponent,
\begin{equation}
z = 3\quad.
\label{eq:6.16b}
\end{equation}
The mean--field critical behaviors of the charge and spin diffusion
coefficients have also been determined\cite{opmit}, with the result
that they both are proportional to the order parameter. For the
exponents $s$ and $s_s$ defined in Table\ \ref{table:6.3} this means
\footnote[27]{The conductivity $\sigma$ is related to the charge diffusion 
coefficient $D_c$ according to the Einstein relation 
$\sigma = D_c\partial n/\partial\mu$, with $\partial n/\partial\mu$
the thermodynamic compressibility. The result quoted for $s$ assumes that
$\partial n/\partial\mu$ has a noncritical contribution.}
\begin{equation}
s = s_s = 1/2\quad
\label{eq:6.16c}
\end{equation}
\end{mathletters}%

We now must ask what the upper critical dimensionality for the problem
is, i.e. the dimension $d_c^+$ above which the above critical behavior is
exact. A superficial inspection of the action suggests $d_c^+=4$.
However, it turns out that under renormalization additional contributions
are created that were not in the bare action, and that lead to
$d_c^+=6$\cite{opmit}. The structure of these additional terms is
reminiscient of the structure encountered in the theory of magnets in
random magnetic fields. This becomes plausible if one recalls that the
random potential in the action couples to the electron density,
see Eq.\ (\ref{eq:3.2c}). If we Fourier transform from imaginary time
to Matsubara frequencies, we obtain a term proportional to
\begin{equation}
\int d{\bf x}\,u({\bf x})\sum_n {\bar\psi}_n({\bf x})\psi_n({\bf x})\quad.
\label{eq:6.17}
\end{equation}
Since the expectation value 
$\langle{\bar\psi}_n({\bf x})\psi_n({\bf x})\rangle$
determines the density of states, this means that the random potential
couples to the order parameter field for the phase transition, just like
a random magnetic field couples to the magnetic order parameter. A
detailed technical analysis\cite{opmit} does indeed confirm this analogy.
Not only is $d_c^+ =6$, but many structures of the theory coincide
with those encountered in the theory of the random--field Ising 
model\cite{RFIsing},
including a quartic coupling constant, $u$, that acts as a dangerous
irrelevant variable even below the upper critical dimension $d_c^+=6$.
An explicit $6-\epsilon$ expansion of the critical
exponents yields also results that, at least to one--loop order, coincide
with those for the random--field Ising model. 

The implications of these results for the critical behavior in $d=3$
are discussed in the next subsection.
For dimensions $d>6$, the mean--field results reviewed above
represent the first example of an Anderson--Mott transition, or 
metal--insulator transition of
disordered interacting electrons, for which the
critical behavior has been determined exactly. Remarkably, this
transition has nothing obvious in common with the Anderson--Mott
transitions that are described by the same model in $d=2+\epsilon$ and
that have been discussed in Sec.\ \ref{subsubsec:VI.B.1} above.
In particular, there is no sign here of the various universality
classes that are present in $d=2+\epsilon$. This is not just the
strong universality that one expects from a mean--field theory, as
it persists in the $6-\epsilon$ expansion. It is therefore likely
that the Anderson--Mott transition described above is of a different
nature than those found in $d=2+\epsilon$. Which description is
closer to what actually happens in $3$--$d$ systems is currently not known.

\subsubsection{Anderson--Mott transitions in three dimensions: Conventional 
                                                           scaling scenario}
\label{subsubsec:VI.B.3}

The similarities between the Anderson--Mott transition in high dimensionalities
and random--field magnets mentioned in the last subsection gives rise to
a scaling description of the former\cite{opmit} 
that is quite different from the one presented in Sec.\ \ref{subsubsec:VI.B.1}.
It is important to point out that for the case of random--field magnets,
this scaling scenario almost certainly is {\em not} correct. However,
there are subtle technical differences between the two systems that
suggest that the situation may be different in the case of the 
Anderson--Mott transition. We therefore present this scenario, which
is a conventional (i.e., power--law) scaling description, here, and
a different, more exotic, possibility in Sec.\ \ref{subsubsec:VI.B.4}
below. Which of these two possibilities provides a better description
of the actual behavior in $d=3$ is not known.

There are two crucial ingredients to a conventional scaling description
of the Anderson--Mott transition based on the order parameter formalism
of Secs.\ \ref{subsubsec:III.C.4} and \ref{subsubsec:VI.B.2}. The
first one is the realization that the electron--electron interaction
is renormalization group irrelevant at this transition.
\footnote[28]{This is true despite the fact that the interaction is necessary
for the transition to occur, see Eq.\ (\ref{eq:6.15a}).}
As a result, the frequency mixing that produced nontrivial dynamical
exponents $z$ in the $2+\epsilon$ expansion, is absent here,
the frequency plays the role of an external field that is conjugate to
the order parameter, and $z$ is not independent. If we denote the
scale dimension of the external field by $y_h$, we have
\begin{equation}
z = y_h = \delta\beta/\nu\quad.
\label{eq:6.18}
\end{equation}
The second important point is the existence of a quartic coupling constant,
$u$, that acts as a dangerous irrelevant variable with a scale dimension of
$[u] = -\theta$. This adds a third independent static exponent, $\theta$,
to the usual two.
The order parameter obeys a scaling or homogeneity relation,
\begin{mathletters}
\label{eqs:6.19}
\begin{equation}
N(t,\Omega,u,\ldots) = b^{(2-d-\eta)/2}\ N(tb^{1/\nu},\Omega b^{\tilde z},
                          u b^{-\theta},\ldots)\quad.
\label{eq:6.19a}
\end{equation}
Here we have denoted by ${\tilde z}$
the scale dimension of $\Omega$ or $T$, $[\Omega] = [T] = {\tilde z}$.
Upon elimination of the dangerous irrelevant variable $u$, ${\tilde z}$
turns into the effective dynamical exponent $z$,
\begin{equation}
N(t,\Omega) = b^{(2+\theta -d-\eta)/2}\ N(tb^{1/\nu},\Omega b^z)\quad.
\label{eq:6.19b}
\end{equation}
\end{mathletters}%
This relates the OP exponent $\beta$ to the three independent exponents
$\nu$, $\eta$, and $\theta$ through the scaling law,
\begin{mathletters}
\begin{equation}
\beta = \frac{\nu}{2}\ (d-\theta -2+\eta)\quad.
\label{eq:6.20a}
\end{equation}
The remaining static exponents are given by the usual scaling laws, with
$d\rightarrow d-\theta$ due to the violation of hyperscaling by the 
dangerous irrelevant variable,
\begin{eqnarray}
\delta&=&(d-\theta + 2-\eta)\nu/2\beta\quad,
\nonumber\\
\gamma&=&\nu(2-\eta)\quad.
\label{eq:6.20b}
\end{eqnarray}
\end{mathletters}%
Now we consider the thermodynamic susceptibilities $\partial n/\partial\mu$,
$\gamma_V$, and $\chi_s$. From general arguments given in 
Ref.\ \onlinecite{opmit} we expect all of them to share the same
critical behavior. Denoting their singular parts collectively by
$\chi_{sing}$, we have
\begin{equation}
\chi_{sing}(t,T) = b^{-d+\theta +z}\ \chi_{sing}(tb^{1/\nu},Tb^z)\quad.
\label{eq:6.21}
\end{equation}
This links the static critical behavior of the thermodynamic susceptibilities,
characterized by the exponent $\kappa$, $\chi_{sing}(t)\sim t^{\kappa}$
(see Table\ \ref{table:6.3}), to that of the OP,
\begin{equation}
\kappa = \beta\quad,
\label{eq:6.22}
\end{equation}
where we have used the scaling laws, Eqs.\ (\ref{eq:6.18}), (\ref{eq:6.20b}).
The fact that all of the thermodynamic sysceptibilities scale like the
order parameter is a consequence of the random--field or static disorder
fluctuations being dominant over the quantum fluctuations.

Let us also consider the scaling behavior of the transport coefficients.
The charge, spin, and heat diffusion coefficients, which we collectively
denote by $D$, all obey the same homogeneity relation,
\begin{equation}
D(t,\Omega) = b^{2-z}\ D(tb^{1/\nu},\Omega b^{z})\quad.
\label{eq:6.23}
\end{equation}
By the definition of the exponents $s_s$ and $s_h$ 
(see Table\ \ref{table:6.3}), this yields
\begin{equation}
s_s = s_h = \beta - \nu\eta\quad,
\label{eq:6.24}
\end{equation}
If the thermodynamic density susceptibility, $\partial n/\partial\mu$,
has a noncritical background contribution, then the conductivity exponent $s$
is also given by Eq.\ (\ref{eq:6.24}). Using Eq.\ (\ref{eq:6.20a}), we find
the following generalization of Wegner's scaling law, Eq.\ (\ref{eq:2.4b})
and Table \ref{table:6.3},
\begin{equation}
s = \frac{\nu}{2}(d-2-\theta-\eta)\quad.
\label{eq:6.25}
\end{equation}

With these results, we can write the homogeneity law for the static electrical
conductivity as
\begin{equation}
\sigma(t,T) = b^{-s/\nu}\,\sigma(tb^{1/\nu}, Tb^z) 
            = t^s\,F_{\sigma}(T/t^{\nu z})\quad,
\label{eq:6.26}
\end{equation}
where the scaling function $F_{\sigma}$ is determined by $\sigma(t=1,T)$, i.e.
by the conductivity far from the transition. The salient point is that,
according to this scaling scenario, $\sigma/t^s$ in the critical region
is a function of $T/t^{\nu z}$, rather than of $T$ and $t$ separately.
This can be checked experimentally. In Eq.\ (\ref{eq:6.26}), the exponent
$s$ is related to the independent exponents $\nu$, $\eta$, and $\theta$
by Eq.\ (\ref{eq:6.25}). An important point in this context is again the
Harris criterion\cite{Harris}, which requires $\nu\geq 2/d$. If Wegner
scaling, Eq.\ (\ref{eq:2.4b}), were valid (as it is, e.g., in the transitions
in $d=2+\epsilon$ discussed in Sec.\ \ref{subsubsec:VI.B.1}), then it would
follow that $\nu\geq 2/3$ in three dimensions. This puts a severe constraint
on any interpretation of experimental results, and has been a much--discussed
issue in the case of the metal--insulator transition that is observed in
Si:P\cite{R}. A thorough discussion of these experiments in the light
of Eq.\ (\ref{eq:6.26}) has been given in Ref.\ \onlinecite{glass}. The
net conclusion was that the data allow for reasonable dynamical scaling
plots, provided that one chooses values of $s$ that are substantially
smaller than $2/3$. An example of such a scaling plot is shown in
Fig.\ \ref{fig:6.3}. 
\begin{figure}[t]
\centerline{\psfig{figure=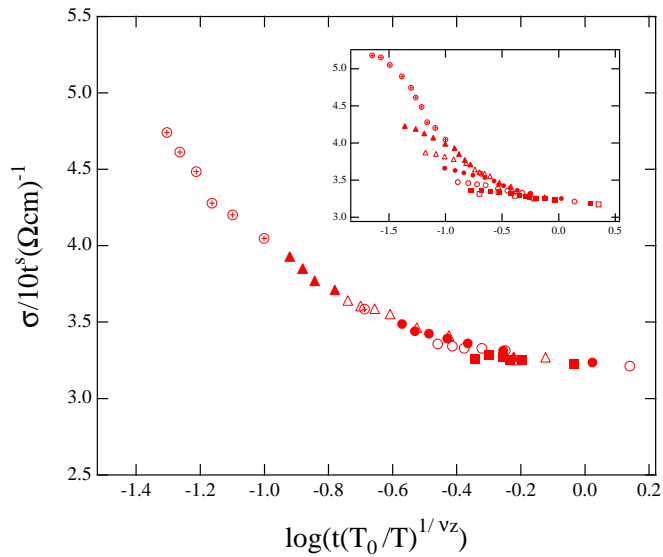,width=90mm}\vspace*{10mm}}
\caption{Dynamical scaling plot of the conductivity data from Fig.\ 1 of
 Ref.\ \protect\onlinecite{BellRC}. The plot assumes a critical stress
 $S_c = 6.71\ {\rm kbar}$, and exponent values $s = 0.29$, $\nu z = 1.82$.
 Only data in the temperature range $T < 60\ {\rm mK}$ have been included
 in the plot, and different symbols denote different stress values, from
 $S=6.59\ {\rm kbar}$ to $S=8.03\ {\rm kbar}$. We have chosen
 $T_0 = 100\ K$, and the relation between $t$
 and $S - S_c$ was taken from Ref.\ \protect\onlinecite{BellLetter1}, viz.
 $t = (S - S_c)\ 5.4 \times 10^{-3}\ ({\rm kbar})^{-1}$.
 The inset shows that the data cease to scale
 once the temperature region $60\ {\rm mK} < T < 225\ {\rm mK}$ is taken into
 account. (From Ref.\ \protect\onlinecite{glass}.)}
\label{fig:6.3}
\bigskip
\end{figure}
This indicated that an interpretation of this
experiment in terms of conventional power law scaling is possible only 
within the framework of a theory that, like the present one, violates
Wegner scaling. This seems to favor the order parameter description of
the Anderson--Mott transition over the $(2+\epsilon)$--dimensional
approach that was reviewed in Sec.\ \ref{subsubsec:VI.B.1}. However,
a competing experiment on the same material\cite{Karlsruhe} obtained
results that are incompatible with Ref.\ \onlinecite{BellRC,BellLetter1},
and allow for
scaling plots of comparable quality with a value of $s\approx 1.3$.
This experimental controversy has not been resolved\cite{BellKCommRep},
and the experimental accuracy is not sufficient for a scaling plot to
discriminate among them. For a more detailed discussion of this point
we refer the reader to Ref.\ \onlinecite{glass}.

\subsubsection{Anderson--Mott transitions in three dimensions: Activated
                                                           scaling scenario}
\label{subsubsec:VI.B.4}

In the previous subsection we gave a scaling theory for the Anderson--Mott
transition, assuming it
was a conventional continuous phase transition, albeit with dangerous
irrelevant variables playing a crucial role. There is, however, another
possibility. We have seen that the Anderson--Mott transition near $d=6$ has
structural similarities to the transition that occurs in a classical
random--field Ising model near $d=6$. The latter has been predicted and
observed to display
glass--like features, and activated rather than power--law scaling in
$d=3$. If the analogy to the Anderson--Mott transition still holds in
$d=3$ (which at this point is merely a subject of speculation), then
the remarkable conclusion is that the Anderson--Mott transition also
has glassy aspects. This possibility has been explored in some detail
in Ref.\ \onlinecite{glass}. Here we briefly show how to modify the
scaling theory for the Anderson--Mott transition in order to allow for
this possibility.

The chief assumption in a scaling theory for a phase transition that shows
activated scaling is that the
critical time scale grows exponentially with $\xi$,
\begin{equation}
\ln \ (\tau /\tau_0)\sim \xi ^\psi \quad ,  
\label{eq:6.27}
\end{equation}
with $\tau_0$ a microscopic time scale, and $\psi$
a generalized dynamical exponent. Physically, this equation implies an
exponential growth of the relaxation time as the transition is approached,
i.e., behavior that is typical of a glass transition. 
As a result of this extreme critical slowing down, the
system's equilibrium behavior near the transition becomes inaccessible for
all practical purposes, and the systems falls out of equilibrium on realizable
experimental time scales. Below we will point out some of the experimental
consequences of this. Before we do so, let us briefly discuss the physical 
ideas behind our picture of a `glassy' Anderson--Mott transition.
As already noted, the Anderson--Mott transition problem contains an intrinsic 
frustration feature: Electron--electron interactions always suppress the 
local density of states, i.e. the
order parameter, while the random potential can cause either local increases or
decreases in the order parameter. This frustration means, for example, that
local insulating clusters exist inside the metallic phase. Elimination of
these clusters requires a large free energy barrier to be overcome. These
barriers are assumed to grow like $L^{\psi}$, with
$L$ some length scale, which near the critical point is given by $\xi$.
Via the Arrhenius law, this leads to Eq.\ (\ref{eq:6.27}).

To construct a scaling theory for this type of transition, the 
usual homogeneity
laws need to be generalized to allow for
activated scaling. Since the barriers are expected to be normally
distributed, while the relaxation times have a much broader distribution,
the natural scaling variables are $\ln\tau /\tau_0$ or $\ln (T_0/T)$, 
with $T_0$ a microscopic temperature scale such as the Fermi
temperature. Considering a variable $Q$, which we assume to be self--averaging,
we therefore expect a homogenity law
\begin{equation}
Q(t,T)=b^{-[Q]}\,F_Q\left( tb^{1/\nu }\,,\,{\frac{b^\psi }{\ln (T_0/T)}}%
\right) \quad ,
\label{eq:6.28}
\end{equation}
where $F_Q$ is a scaling function. For the tunneling density of states 
this implies,
\begin{mathletters}
\label{eqs:6.29}
\begin{equation}
N(t,T)=b^{-\beta/\nu}F_N\left( tb^{1/\nu }\,,\,{\frac{b^\psi}{\ln (T_0/T)}}
                                                             \right) \quad .
\label{eq:6.29a}
\end{equation}
Eliminating the parameter $b$ by choosing $b^{\psi} = \ln (T_0/T)$ gives
\begin{equation}
N(t,T)={\frac 1{\left[ \ln (T_0/T)\right] ^{\beta /\nu \psi }}}\ G_N\left[
t^{\nu \psi }\ln (T_0/T)\right] \quad ,
\label{eq:6.29b}
\end{equation}
\end{mathletters}%
with $G_N$ another scaling function. Equation (\ref{eq:6.29b}) illustrates
a general and important result:
At criticality, $t=0$, the critical singularities will be only 
logarithmic in
$\ln (T_0/T)$. This is the quantum analog of the fact that
classical static glass transitions are experimentally inacessible.

Other quantities that are singular at the critical point have been discussed
in Ref. \onlinecite{glass}. Here we explicitly consider the magnetic 
susceptibility, and the
specific heat coefficient. Given that the entropy and the magnetization,
as thermodynamic quantities, satisfy 
homogeneity laws like Eq.\ (\ref{eq:6.28} ), it has been argued\cite{glass} 
that neither $\chi_m$ nor $\gamma_V$ satisfy such a law. As a
consequence of this, one finds that these quantities are singular even
in the metallic phase. For example, $\chi_m$ scales like
\begin{equation}
\chi_m(t,T) = {\frac{T^{\,-1\,+\,{\rm const}\times
                                 t^{\nu\psi}}}{\left[\ln(T_0/T)%
                    \right]^{(d-\theta)/\psi - 2\phi}}}\quad.
\label{eq:6.30}
\end{equation}
$\phi$ is an exponent related to the difference in scaling between $T$ and
a magnetic field $H$. The singular behavior in the metallic phase means
that there is a Griffiths phase, or a region away from the critical point in
which certain observables become singular at various values of $t$.

Finally, it has been argued in Ref.\ \onlinecite{glass} that the 
conductivity, $\sigma$, is
not self averaging and that instead one should consider 
$\ell_{\sigma} \equiv \left\{\ln (\sigma_0/\sigma)\right\}_{dis}$,
with $\sigma_0$ the bare conductivity. The physically reason for this is 
that $\sigma$ is related to a
relaxation time which, as mentioned above, is log--normally distributed.
$\ell_{\sigma}$ obeys a scaling law
\begin{eqnarray}
l_\sigma (t,T) &=&b^\psi F_\sigma \left( tb^{1/\nu }\,,\,{\frac{b^\psi }{\ln
(T_0/T)}}\right) \quad \quad \ \ \,  \nonumber \\
&=&\ln (T_0/T)\ G_\sigma \left( t^{\nu \psi }\ln (T_0/T)\right) \quad .
\label{eq:6.31}
\end{eqnarray}
The important point is that the observable conductivity at $T=0$ will
vanishes exponentially fast as $t\rightarrow 0$, but that it is impossible 
to observe this behavior because it requires
exponentially low temperatutes. The same arguments also imply that
one should expect large sample--to--sample variations in the measured
conductivity.

There is some experimental support for the glassy picture presented above.
First, it has been known for some time that there are large sample--to--sample
variations in the conductivity measurements in doped semiconductors. Already
in the early 1980's this was observed in Si:P for $t\leq 10^{-3}$ in the
mK temperature range\cite{Rosenbaum,BellLetter1}. 
Later, similar effects were blamed on some
type of thermal decoupling\cite{Karlsruhe}. 
The so--called Griffiths like phase in which
$\chi_m$ and $\gamma_V$ are singular already in
the metallic phase has also been seen in the doped semiconductors.
Coventionally this behavior has been attributed to local moment 
formation\cite{Paalanenetal}, and the vanishing of the Kondo 
temperature\cite{BhattFisher,Dobroetal} in disordered systems. Whether,
or how, these
effects are related to the Griffiths singularities that arise in our glassy
phase transition theory is not obvious and warrants further work.

\acknowledgments
We would like to thank our collaborators on some of the work on magnetic
phase transitions reviewed above, Thomas Vojta and Rajesh Narayanan.
This work was supported by the NSF under grant numbers DMR--96--32978 and
DMR--95--10185.

\vfill\eject
\end{document}